\documentclass[a4paper,11pt]{article}
\pdfoutput=1 % if your are submitting a pdflatex (i.e. if you have
\usepackage{jcappub} % for details on the use of the package, please
\usepackage[T1]{fontenc} % if needed

\usepackage{soul,xcolor}
\usepackage{tikz}
\usepackage{xparse}
\definecolor{dred}{rgb}{0.7, 0.0, 0.0}

\newcommand{\highlight}{}
\NewDocumentCommand{\xarrows}{ O{}O{} }{%
\mathrel{%
\vcenter{\hbox{%
\begin{tikzpicture}
  \node[minimum width=1cm,minimum height=1ex,anchor=south,align=center] (a){\text{\vphantom{hg}#1}\\[0.5ex] \vphantom{hg}#2};
  \draw[->] ([yshift=0.35ex]a.west) -- ([yshift=0.35ex]a.east);
  \draw[<-] ([yshift=-0.35ex]a.west) -- ([yshift=-0.35ex]a.east);
\end{tikzpicture}
}}%
}%
}

\def\MNRAS{\emph{Mon. Not. Roy. Astron. Soc. }}

\title{\boldmath Cored DARKexp systems with finite size: numerical results}

\author{Claudio Destri}
\affiliation{Dipartimento di Fisica G. Occhialini, Universit\`a
Milano-Bicocca\\ and INFN, sezione di Milano-Bicocca, Piazza della Scienza 3,
20126 Milano, Italia.}
\emailAdd{claudio.destri@mib.infn.it}

\abstract{In the DARKexp framework for collisionless isotropic relaxation of self--gravitating
  matter, the central object is the differential energy distribution $n(E)$, which takes a
  maximum--entropy form proportional to $\exp[-\beta(E - \Phi(0))] - 1$, $\Phi(0)$ being the depth
  of the potential well and $\beta$ the standard Lagrange multiplier. Then the first and quite
  non--trivial problem consists in the determination of an ergodic phase--space distribution which
  reproduces this $n(E)$. In this work we present a very extensive and accurate numerical solution of
  such DARKexp problem for systems with cored mass density and finite size. This solution holds
  throughout the energy interval $\Phi(0)\le E\le 0$ and is double--valued for a certain interval of
  $\beta$. The size of the system represents a unique identifier for each member of this solution
  family and diverges as $\beta$ approaches a specific value. In this limit, the tail of the mass
  density $\rho(r)$ dies off as $r^{-4}$, while at small radii it always starts off linearly in $r$,
  that is $\rho(r)-\rho(0)\propto r$.}

\begin{document}
\maketitle
\flushbottom

\section{Introduction and summary}\label{sec:intro}

One of the long--lasting question in statistical physics concerns the equilibrium, or
\emph{quasi}--equilibrium in collisionless self--gravitating systems. In the limit of continuous
matter, the dynamics in such systems is well described in the single particle phase--space by the
Vlasov--Poisson (or collisionless Boltzmann) equation \cite{braun1977,chavanis2008}. But this
equation retains full memory of the initial conditions at the fine--grained level. Yet observations
and $N-$body simulations show that collisionless systems of many different kinds and sizes appear,
at the coarse--grained level, to be in very similar quasi--stationary states
\cite{bt,padmanabhan1990,aarseth2003,navarro2010}.

This fact naturally suggests that some sort of maximum--entropy principle should be at work also for
the outcome of collisionless gravitational collapses, and many proposals in this direction have been
put forward throughout the years
(\cite{lyndenbell1967,stiavelli1987,white1987,hjorth1991,spergel1992,plastino1993,chavanis1998,hjorth2010,he2010,pontzen2013}
is a very incomplete list), although no general framework or specific model has yet obtained the
success and recognition that standard Gibbs--Maxwell--Boltzmann equilibrium enjoys for systems with
short--range interactions.

\smallskip 
Apart from well--known intrinsic theoretical difficulties, one reason for this unsatisfactory state
of the art lays in the many limitations and subtleties that dominate the scene when observations are
compared with $N-$body simulations and both are contrasted with first--principle theoretical
models. First of all, astrophysical observations have a well--known limited discriminating power and
very often the data themselves depend on more or less arbitrary assumptions on the underlying
dynamics.  Secondly, even when we are convinced to be dealing with purely self--gravitating matter
in a collisionless regime, such as with dark matter, it is not easy to disentangle a localized
gravitational collapse from all the rest. In dark--matter--driven cosmological structure formation
there are hierarchical collapses, filaments, mergers, tidal strippings and other complex effects
that make it difficult to unambiguously identify relaxed outcomes of well defined collapses. When
eventually well--localized quasi--stationary systems emerge, such as galaxies, baryons with their
complicated non--gravitational interactions could play a role also in the shaping of the inner part
of the dark matter halo that hosts a visible galaxy
\cite{navarro1996,governato2012,pontzen2012,chan2015,brooks2017}. Furthermore, given that we only
have an indirect knowledge of DM, we cannot apriori exclude the possibility of a self--interacting
DM whose elastic collisions could significantly alter the collisionless density profiles observed in
$N-$body simulations
\cite{spergel2000,vogelsberger2012,rocha2013,zavala2013,vogelsberger2016}. Another possibility is to
vary the initial conditions of structure formation and in particular the dispersion velocity of the
hypothetical DM particles; this could very well affect also the outcomes of localized small--scale
collapses. Cold DM suffers from the so--called small--scale crisis, as it yields too much structure
at small scales and cuspy density profiles \`a la Navarro--Frenk--White. Warm DM avoids the small
scale overabundance \cite{colombi1996,bode2001} (but perhaps entering in conflict with
Lyman$-\alpha$ observation data \cite{viel2013}) and is expected to yield cored profiles
\cite{hogan2000,boyanovsky2008,boyarski2009,devega2010,maccio2013}, although probably too small to
reconcile $N-$body simulations with observations without the help of baryonic feed--back
\cite{villaes2011,maccio2013,lovell2014}. If DM is warm, even Quantum Mechanics might play a role in
the quasi--equilibrium of small dwarf galaxies \cite{destri2013a,destri2013b}, which are the systems
where observations and pure CDM $N-$body simulations disagree the most. And certainly this is only a
partial list of possibilities.

For this reason, in our opinion a practical and effective approach to the theoretical problem should
start by clearly delimiting the scope of the investigation, leaving possible applications, with all
the necessary provisos and adjustments, to a second stage, but dealing with the theory, whether
analytically or numerically, as carefully as possible. This is the framework we choose for the
present work.

\smallskip
There are two fundamental prerequisites that any maximum--entropy model of the collisionless
collapse ought to satisfy: the system should be in dynamical equilibrium, that is the phase--space
mass distribution should be a function of the isolating integrals of the motion, and it should have
finite mass and energy. One such model is the DARKexp model, originally put forward by Hjorth and
Williams in \cite{hjorth2010} with the explicit purpose of providing a statistical--mechanical basis
for the Navarro--Frenk--White density profile \cite{nfw}. In its simplest version, which applies to
fully isotropic systems, the model envisage an ergodic phase--space mass distribution $f$ that
depends only on the orbit energy $E=\frac12 v^2+\Phi$, with $\Phi(r)$ the spherically symmetric
gravitational potential, just as many other well--known models \cite{bt}. The crucial difference
w.r.t. to other maximum--entropy formulations, such as the isothermal sphere, is that the
maximization of Boltzmann--Shannon entropy is assumed in energy space rather than in phase space
(see \cite{hjorth2010} for the motivations). Moreover, the small occupation numbers of energy states
close to the bottom of the potential well is estimated more accurately than with the Stirling
approximation, as done for instance in finite--mass collisional systems near the escape energy
\cite{madsen1996}. It follows that only the differential energy distribution $n(E)$ \cite{bt} is
explicitly derived in the DARKexp model and reads
\begin{equation}\label{eq:nofe}
  n(E) \propto \exp[-\beta(E - \Phi(0))] - 1 \;,
\end{equation}
where $\beta$ is a Lagrange multiplier that fixes the mean energy of the system, while the overall
normalization is fixed by the total mass. Clearly, by construction, a DARKexp systems has a finite
mass and a finite energy. The phase--space mass distribution $f(E)$, and with it all physical
observables such as the mass density $\rho(r)$ and the velocity dispersion $\sigma(r)$, must be
computed from $n(E)$. This is a highly non--trivial task, which is formulated in detail in section
\ref{sec:dless}, since the $n(E)$ to $f(E)$ relation depends on the potential $\Phi(r)$, that is not
known apriori. Eventually, the reconstructed DARKexp $\rho(r)$ and $\Phi(r)$ depend on three free
parameters: the two arbitrary scales in mass and distances (the third scale is fixed by Newton's
constant) and a dimensionless shape parameter related to $\beta$. {\highlight A fundamental analytic
  property of the DARKexp setup was highlighted in \cite{hjorth2010}: if $\rho(r)$ has the a $1/r$
  cusp as $r\to0$, then $f(E) \sim (E - \Phi(0))^{-5/2}$, $\Phi(r)-\phi(0) \sim r$ and $g(E)\sim (E -
  \Phi(0))^{7/2}$ as $E\to \Phi(0)$, so that $n(E) \sim E - \Phi(0)$ in agreement with
  eq.~\eqref{eq:nofe}.}

\medskip {\highlight Strictly speaking, the DARKexp framework just introduced has a logical loophole
  if it is meant solely as a statistical--mechanical foundation to NFW $1/r$ profile, the prototype
  of the cuspy profiles characteristic of pure CDM $N-$body simulations. In fact, DARKexp deals with
  continuous matter and therefore it applies to the realistic collapse of DM particles of arbitrary
  mass (as long as their number in any resolvable volume is large enough to neglect the collisional
  dissipation). And yet, it is known that the primordial DM velocity dispersion, which strongly
  depends on the DM particle mass, has a direct effect on the inner profile of relaxed halos. In
  other words, DARKexp whould predict $1/r$ cusps also for relaxed WDM, which instead features
  cores, albeit apparently not large enough to agree with observations. Therefore, it is reasonable
  to investigate also the possibility of DARKexp cored systems and to justify the use of the DARKexp
  model more on its acceptable consistency with observations and $N-$body simulation than on its
  implication of $1/r$ cusps.}

\smallskip
{\highlight Indeed,} the DARKexp model has been tested against observations and/or $N-$body
simulations in two distinct ways: in \cite{williams2010b,nolting2016} the $n(E)$ of
eq.~\eqref{eq:nofe} was compared to the differential energy distribution extracted from $N-$body
simulations. In \cite{beraldo2013,umetzu2016,nolting2016} the mass density implied by $n(E)$ (and
numerically computed) was directly compared to observations and/or $N-$body simulations. In both
cases the DARKexp model fares quite well. However, in both cases there are obvious limitations to
the finite resolving power, either for energies very close to the potential depth or for very small
spatial distances from the center of mass. Similarly, there are limitations near the escape energy
or at large distances. The situation becomes even more involved in the case of the DARKexp
$\rho(r)$, since it has to be computed numerically from $n(E)$ and it may suffer itself from
limitations at small and large distances. This is the case for the numerical method just mentioned
in \cite{williams2010a} and briefly described on \cite{beraldo2013}, which is originally due to
Binney \cite{binney1982} and is affected by numerical instabilities at distances too small or to
large. Moreover, this recursive method might very well suffer also from numerical artifacts even
where it appears to be stable, especially if the mass density $\rho(r)$ blows up when $r\to0$ as in
the case of the NFW profile. The density profiles and their logarithmic slopes plotted in Figs.~1
and 2 of \cite{williams2010a} do show peculiar oscillations that give no hint to settle on the
$r\to0$ asymptotics of NFW profile, which is nonetheless claimed to be a consequence of the DARKexp
model. Apparently, this is not a problem for the good fit to observations and $N-$body simulations,
since this is confined to a region where the oscillations are negligible, but it does make the basis
of the DARKexp model less firm than it is reasonable to accept.

\medskip 
With the motivation of clarifying the $n(E)$ to $\rho(r)$ relation of the DARKexp model, in
\cite{destri2018} we performed a next--to--leading asymptotic analysis when the leading behavior of
$\rho(r)$ is assumed to be the cuspy one of the NFW profile or just a standard core like in most
ergodic systems. In fact, contrary to some common lore, in \cite{destri2018} it was shown that cored
profiles are perfectly consistent with a $n(E)$ linearly vanishing when $E\to\Phi(0)$ as in
eq.~\eqref{eq:nofe}. Furthermore, it was shown that the next--to--leading asymptotics in the case of
cores is much better behaved that in the case of $1/r-$cusps. In the present work we complete the
analysis in the case of cored systems with a finite size by computing very accurate numerical
solutions of the DARKexp problem that perfectly match the asymptotic behavior analytically
determined in \cite{destri2018}. In other words, we obtain the profiles for $f(E)$ and $\rho(r)$ at many
different values of $\beta$ with no limitations on the range of $E$ or $r$. In particular,
this means that our numerical solutions confirm the behavior of $\rho(r)$ near $r=0$ reported in
eq.~7.2 of \cite{destri2018}, namely
\begin{equation*}
  \rho(r) \simeq \rho_0\left[1 + \gamma_0r - 
    \frac{\!\!32}{5\pi^2}\,\gamma_0^2\,r^2\log r + O(r^2)\right]\;,
\end{equation*}
where $\rho_0$ and $\gamma_0$ represent the two free scales of the problem. 

\smallskip 
Our numerical approach is described in detail in section \ref{sec:numapp}. It is based on the
precise dimensionless formulation of the DARKexp problem put forward in \cite{destri2018} and
repeated here in section \ref{sec:dless} for completeness. This formulation lends itself very
naturally to the use of the Chebyshev approximation as central numerical tool. The results are
described in detail in section \ref{sec:results}. A quick summary is provided by
Figs.~\ref{fig:rho_slopes} and~\ref{fig:sigmassq}, where we plot the mass density profiles, their
logarithmic slopes and the velocity dispersions. The density and its logarithmic slope are always
monotonic. For obvious reasons in these figures the logarithm of radial coordinate is bound on the
left, but our numerical approach is capable of computing $\rho(r)$ with $O(10^{-10})$ accuracy for
any value of $r$, no matter how small.

Another prominent and somewhat unexpected feature is the dependence of the DARKexp solutions on the
Lagrange multiplier $\beta$, which plays a role similar to that of a negative inverse temperature.
This behavior turns out to be double--valued, with an upper limit on $\beta$ in one of the two
determinations, as summarized in Fig.~\ref{fig:Kmxeta_vsb} ($b$ is a dimensionless variable
proportional to $\beta$).

Altogether, the cored DARKexp systems with finite size that we have computed represent an
interesting family of ergodic self-gravitating systems, regardless of their probably limited
applicability to actual physical situations. They range from very compact and hot, almost uniform
spheres, to cold and diffused halos with a $r^{-4}$ tail that eventually drops to zero arbitrarily
far away from the origin, since the size of the system diverges when $\beta$ tends to a specific
limiting value. Let us recall that we are dealing here with continuous ergodic systems that are
completely relaxed in energy. In real gravitational collapses, for instance of DM, even assuming
perfect spherical symmetry there remains the magnitude of angular momentum as isolating integral of
the motion, which is connected to the velocity anisotropy. Indeed the DM halos in $N-$body
simulations have a characteristic non--zero anisotropy $\beta(r)$. In \cite{williams2010b} it was
argued that if DARKexp halos had similar $\beta(r)$, their density and energy distributions could
not be distinguished from those of isotropic DARKexp halos. This line of argument was further
developed in \cite{williams2014}, where the role of angular momentum in the DARKexp maximum--entropy
setup was carefully studied with a mixed theoretical and phenomenological approach. In both cases
the DARKexp $n(E)$ was regarded as implying a $1/r-$cusped density with effectively unlimited
support. It should be intersting to extend this kind of analysis to the cored DARKexp systems with
finite size presented here.

\section{Dimensionless formulation for cored DARKexp systems} \label{sec:dless} 

A spherically symmetric system is characterized by the Poisson's equation
\begin{equation}\label{eq:pois0}
  \phi''+\frac2r \phi' = 4\pi G\rho\;.
\end{equation}
Here we assume that the mass density $\rho=\rho(r)$ is finite at $r=0$, yields a finite total mass
$M$ and vanishes when $r\ge r_0$ for some $r_0$, possibly infinite, that defines the size of the
system. We then fix the arbitrary constant mode of the gravitational potential $\phi$ by demanding
that $\phi(r_0)=0$. By Gauss law $\phi$ is monotonically increasing in $r$ and so $\phi(r)<0(>0)$
for $r<r_0(>r_0)$ with $\lim_{r\to\infty}\phi(r)=G M/r_0$.

\smallskip
Choosing units such that $4\pi G = 1$, there remain two arbitrary scales for eq.~\eqref{eq:pois0},
which can be fixed in terms of $\rho_0=\rho(0)$ and $\phi_0=\phi(0)$. Hence we can set
\begin{equation*}
   \rho(r) = \tfrac13\rho_0\, \nu(u(x))\;,\quad \phi(r)=\phi_0[1-u(x)]
   \;,\quad x\equiv\frac{r}{r_\ast} \;,\quad r_\ast^2 = \frac{3|\phi_0|}{\rho_0}\;,
\end{equation*} 
where $\nu(u)$ and $u(x)$ are dimensionless functions of their own dimensionless argument and $u(x)$
monotonically grows from $0$ to $1$ as $x$ grows from $0$ to $r_0/r_\ast$. By construction
$\nu(0)=3$ and Poisson's equation now reads
\begin{equation}\label{eq:pois1}
   u''(x) +\frac2x u'(x) = \nu(u(x)) \;.
\end{equation}
while the solution must behave as $u \simeq \frac12 x^2$ when $x\to0$.  

\smallskip
The monotonic relation between $x$ and $u$ also allows to rewrite eq.~\eqref{eq:pois1} as
\begin{equation}\label{eq:pois2}
   \frac{2}{x(u) x'(u)} - \frac{x''(u)}{[x'(u)]^3} = \nu(u) \;,
\end{equation}
with $x \simeq (2u)^{1/2}$ as $u\to0$. Notice that in this last reformulation of Poisson's equation
the non--linearity is fixed independently of the form of $\nu(u)$ and the domain of the now
independent variable $u$ is fixed once and for all to $[0,1]$, with the system bounded whenever
$x(1)=r_0/r_\ast$ is finite. On the other hand, eq.~\eqref{eq:pois2} hides an important positivity
property of Poisson's equation which is instead manifest in eq.~\eqref{eq:pois1}, since this is
equivalent to the integral equation
\begin{equation}\label{eq:intpois}
  u(x) = \int_0^xdy\,y^{-2}\!\int_0^ydz\,z^2\nu(u(z))= \int_0^xdy\,y^2(y^{-1}-x^{-1})\nu(u(y))\;.
\end{equation} 
We can see that any variation of $\nu(u)$ with a definite sign implies a variation of $u(x)$ with
the same sign and therefore a variation of $x(u)$ with the opposite sign.

\medskip 
In its simplest version, the DARKexp model describes isotropic systems with an ergodic phase--space
mass distribution $f=f(\frac12 v^2+\phi(r))$, from which $\rho$ is recovered as
\begin{equation}\label{eq:f2rho}
  \begin{split}
    \rho(r) &= \int\!d^3v\,f(\tfrac12 v^2+\phi(r)) = \int\!dE\,f(E)\int d^3v\, 
    \delta(\tfrac12 v^2+\phi(r)-E)\\
    &= 4\pi \int_{\phi(r)}^0 \!dE\,f(E)\,[2(E-\phi(r))]^{1/2}\;.
  \end{split}
\end{equation}
Here we have used that $f(E)=0$ for $E>0$ as dictated by the condition of bound matter with finite
total mass. If we parametrize the one--particle energy as the potential, that is 
\begin{equation}\label{eq:s2E}
  E=\phi_0(1-s) \;,\quad 0\le s\le 1 \;,
\end{equation}
then we can write
\begin{equation*}
  f(E) = \frac{(2|\phi_0|)^{-1/2}}{4\pi\,r_\ast^2}\,F(s) \;,
\end{equation*}
with $F(s)$ dimensionless and
\begin{equation}\label{eq:F2nu}
   \nu(u) = \int_u^1\!ds\, F(s)\,(s-u)^{1/2} \;.
\end{equation}
This integral transform is the half--primitive of $F(s)$ vanishing at $s=1$ and can be inverted into
the half--derivative
\begin{equation}\label{eq:nu2F}
  F(s) = \frac{2}{\pi}\frac{d}{ds}\int_s^1\!du\, \nu'\!(u)\,(u-s)^{-1/2}\;.
\end{equation}
The differential energy distribution \cite{bt} reads
\begin{equation*}
   n(E) = \frac{dM}{dE} = \int\!d^3r\!\int\!d^3v\,f(\tfrac12 v^2+\phi(r)) 
   \delta(\tfrac12 v^2+\phi(r)-E) = f(E)g(E) \;,
\end{equation*}
where $g(E)$ is the density of states at a given one--particle energy $E$,
\begin{equation}\label{eq:g}
    g(E) = \int\!d^3r\!\int\!d^3v\,\,\delta(\tfrac12 v^2+\phi(r)-E) =
    (4\pi)^2 \int_0^{r_E} \!dr\,r^2\,[2(E-\phi(r))]^{1/2}
\end{equation}
and $r_E$ is the radius where the one--particle kinetic energy vanishes, that is $E=\phi(r_E)$.  By
definition we have
\begin{equation}\label{eq:M}
  M = \int_{\phi_0}^0\!dE\,n(E) \;.
\end{equation}
In dimensionless terms
\begin{equation}\label{eq:M2m}
  M = 4\pi r_\ast|\phi_0|\,m =  \frac{4\pi}3\rho_0r_\ast^3m \;,\quad 
  m = \int_0^1du\, x^2(u)x'(u)\nu(u) 
\end{equation}
and
\begin{equation*}
    n(E) = 4\pi r_\ast N(s) \;,\quad g(E) = (4\pi)^2\,r_\ast^3\,(2|\phi_0|)^{1/2} \,G(s) \;,
\end{equation*}
where
\begin{equation*}
   \quad N(s) = F(s)G(s) \;,\quad  \int_0^1\!ds\,N(s) = m
\end{equation*}
and
\begin{equation}\label{eq:x2G}
   G(s) = \int_0^s\!du\,x^2(u)x'(u)\,(s-u)^{1/2} = \frac16 \int_0^s\!du\,x^3(u)\,(s-u)^{-1/2}\;,
\end{equation}
This integral transform (the half--primitive of $x^2(u)x'(u)$ vanishing at $s=0$) can be inverted as
\begin{equation}\label{eq:G2x}
  x^3(u) = \frac{6}{\pi}\int_0^u\!ds\, G'(s)\,(u-s)^{-1/2} \;.
\end{equation}
Altogether the four functions $F$, $\nu$, $x$ and $G$ are connected as follows
\begin{equation}\label{eq:chain}
  F(s)\, \xarrows[\small eq.~\eqref{eq:F2nu}][\small eq.~\eqref{eq:nu2F}]\,\nu(u)
  \,\xarrows[\small eq.~\eqref{eq:pois2}][\small eq.~\eqref{eq:pois2}] \,x(u)\,
    \xarrows[\small eq.~\eqref{eq:x2G}][\small eq.~\eqref{eq:G2x}] \,G(s) \;.
\end{equation}
and the central problem of any approach in which $n(E)$ is given beforehand, is to find functions $F$
and $G$ such that $F(s)G(s)=N(s)$, taking into account that by eq.~\eqref{eq:chain} $F$ and $G$ are
complicated non--local and non--linear functionals one of the other. In particular, the DARKexp 
differential energy distribution has the form
\begin{equation*}
  n(E) = C\, \{\exp[-\beta(E - \phi_0)] - 1\} \quad,\qquad \phi_0 \le E\le 0 \;,
\end{equation*}
where the two dimension--full constants $C$ and $\beta$ act as Lagrange multipliers that fix the total
mass and the average one--particle energy, respectively. In the present dimensionless setup, we set
\begin{equation*}
  b = \beta \phi_0 \;,\qquad C' = \frac{C}{4\pi\,r_\ast} 
\end{equation*}
so that 
\begin{equation}\label{eq:FG}
  N(s) = F(s)G(s) = C'\, (e^{bs} - 1) = \frac{mb}{e^b-1-b}\,(e^{bs} - 1) \;,
\end{equation}
where the last equality, fixing $C'$, ensures that $\int_0^1\!ds\,N(s)=m$, while the dimensionless
parameter $b$ relates to the average value of the one--particle energy, $\bar{E}=\phi_0(1-\bar{s})$,
through
\begin{equation}\label{eq:bars}
  \bar{s} = \frac{\int_0^1\!ds\,s(e^{bs}-1)}{\int_0^1\!ds\,(e^{bs}-1)} = 
  \frac{(b-1)e^b+1-b^2/2}{b(e^b-1-b)}\;.
\end{equation}
$N(s)$ is positive--defined, as necessary, both for $b<0$ and $b>0$. In both cases it is also
monotonically increasing in $s$, that is in the one--particle energy $E$. On the other hand, $N(s)$
is convex for $b>0$ and concave for $b<0$, while it reduces to a linear ramp for $b=0$.  As $b$
ranges from $-\infty$ to $+\infty$, $\bar s$ monotonically grows from $1/2$ to $1$, so that a
solution for $b$ exists only for $\bar{s}>1/2$ and is unique. Moreover, $b(\bar{s})$ is negative for
$1/2<\bar{s}<2/3$ and positive for $\bar{s}>2/3$.

\medskip
Eq.~\eqref{eq:FG} implies that $F(s)G(s) \sim s$ as $s\to0$. Since $x \simeq (2u)^{1/2}$ as $u\to0$,
we have 
\begin{equation}\label{eq:asyG}
    G(s) \simeq \tfrac{1}{4\sqrt2}\pi s^2 \;,
\end{equation}
where we used the definition of Euler's Beta function
\begin{equation*}
  \int_0^s\!du\,u^{z-1}(s-u)^{w-1} = B(z,w)s^{z+w-1} 
\end{equation*}
along with $B(\tfrac32,\tfrac32)=\tfrac18\pi$. Hence
\begin{equation}\label{eq:asyF}
  F(s) \simeq K s^{-1} \;,
\end{equation}
for some constant $K$ to be determined. When $u=0$, the $s^{-1/2}$ singularity in the integrand of
eq.~\eqref{eq:F2nu} is integrable, hence all values of $F(s)$ over the range $0\le s\le 1$
contribute to $\nu(0)=3$ and $K$ cannot be determined without solving the full problem. To this end,
it is convenient to set
\begin{equation}\label{eq:F2xi} 
  F(s) = Ks^{-1}{\highlight (1-h s)}^{5/2}\xi_F(s)
\end{equation}
and
\begin{equation}\label{eq:G2xi} 
  G(s) = 2^{-5/2}\pi s^2{\highlight(1-h s)}^{-5/2}\xi_G(s) \;, \\[1ex]
\end{equation}
where $h$ is a free parameter, with $0\le h\le1$. The reason for this peculiar parametrization will
become clear later on. Now the problem $F(s)G(s)=N(s)$ takes the form 
\begin{equation}\label{eq:xixi}
   \xi_F(s)\xi_G(s) = A(bs) \;,\quad A(z) \equiv \frac{e^z-1}{z}
\end{equation}
while
\begin{equation}\label{eq:Km}
    K = \frac3{\int_0^1\!ds\,{\highlight (1-h s)}^{5/2}\xi_F(s)s^{-1/2}}\;,
    \quad m = 2^{-5/2}\pi K\frac{e^b-1-b}{b^2}\;;\\[2ex]
\end{equation}
A more comprehensive notation, highlighting all dependencies on the dimensionless variables $s$, $b$
and $h$ as well as the functional dependence of $G$ on $F$ through the chain 
\eqref{eq:chain}, would be
\begin{equation*}
    F(s,b) = K(b)s^{-1}{\highlight (1-h s)}^{5/2}\xi_F(s,b,h) \;,
\end{equation*}
\begin{equation*}
    G(s,b;[F]) = 2^{-5/2}\pi s^2{\highlight (1-h s)}^{-5/2}\xi_G(s,b,h,[\xi_F]) \;, \\[1ex]
\end{equation*}
and
\begin{equation*}
   \xi_F(s,b,h)\xi_G(s,b,h,[\xi_F]) = A(bs) \;,
\end{equation*}
but we will stick to the shorter notation used before, for the sake of brevity. It should be noted,
however, that $\xi_F$ and $\xi_G$, and therefore also $K$ and $m$, could multi--valued as functions
of $b$ in case the solution of eq.~\eqref{eq:xixi} is not unique. Notice also that, by construction,
$\xi_F(0)\xi_G(0) = A(0)$ identically.

\subsection{The case of infinite size}\label{sec:infinite}

The size of the system is given by $r_\ast x(1)$ and it could be infinite. That is to say, for
certain values of $b$ the solution of eq.~\eqref{eq:xixi} might be such that $\xi_F(1)=0$ and
$\lim_{s\to1}\xi_x (1) = \lim_{s\to1}\xi_G(s)=+\infty$. In fact, a finite mass system always has a
Newtonian potential dying off at infinity as $GM/r$. Hence, if $\lim_{u\to1}x(u)=+\infty$, it must
be as a pole of order one, since
\begin{equation*}
  \lim_{u\to1}(1-u)x(u) = m\;.
\end{equation*} 
Then, by eq.~\eqref{eq:x2G}, $G(s)$ must diverge at $s=1$ as $(1-s)^{-5/2}$ {\highlight [see few
lines below eq.~\eqref{eq:IxG} for a rigorous derivation]} , which in turn implies
that $F(s)$ must vanish there as $(1-s)^{5/2}$ [so that $\nu(u)$ dies as $(1-u)^4$ at $u=1$, or
$\rho(r)\sim r^{-4}$ as $r\to\infty$]. This fact explain our peculiar parametrization in
eqs.~\eqref{eq:F2xi} and~\eqref{eq:G2xi}, which allows, by letting $h\to1$, to keep
$\xi_G(s)$ a bounded function over the unit interval $[0,1]$, which is much easier to handle
numerically. As a matter of fact, this parametrization is numerically very convenient 
also for systems with finite but very large size, since $h$ can be tuned to keep $\xi_G(1)$
relatively small when $G(1)$ becomes very large.

\subsection{About energies}

A system described by the ergodic phase--space mass distribution $f(E)$ is isotropic, with 
the squared velocity dispersion given by
\begin{equation}\label{eq:sigma}
  \sigma^2(r) = \frac1{\rho(r)}\int\!d^3v\,v^2\,f(\tfrac12 v^2+\phi(r)) =
  \frac{4\pi}{\rho(r)} \int_{\phi(r)}^0 \!dE\,f(E)\,[2(E-\phi(r))]^{3/2}\;.  
\end{equation}
Hence the total kinetic energy reads
\begin{equation*}
  {\cal K} = \frac12  \int\!d^3r \rho(r)\sigma^2(r) = \frac12 \int_{\phi_0}^0 \!dE\,f(E)g_3(E) \;,
\end{equation*}
where (recall that $E=\phi(r_E)$)
\begin{equation}\label{eq:g}
    g_3(E) = (4\pi)^2 \int_0^{r_E} \!dr\,r^2\,[2(E-\phi(r))]^{3/2} = 
     (4\pi)^2 \int_0^{r_E} \!dr\,r^3\phi'(r)\,[2(E-\phi(r))]^{1/2} \;.
\end{equation}
The total potential energy reads
\begin{equation*}
  {\cal U} = -\int\!d^3r\,\rho\,\boldsymbol{r}\cdot\nabla\phi = 
  -\,4\pi\!\int_0^{r_0}\!dr\,r^3\,\rho(r)\phi'(r) 
  = -\int_{\phi_0}^0 \!dE\,f(E)g_3(E) \;,
\end{equation*}
verifying the well--known virial theorem ${\cal U} = -2{\cal K}$, so that the total energy is
\begin{equation*}
  {\cal E} = {\cal K} + {\cal U} = \tfrac12 {\cal U} = -{\cal K} \;.
\end{equation*}
The total energy is connected to the average one--particle energy, since 
\begin{equation*}
  \begin{split}
    M\bar E &= \int_{\phi(0)}^0 \!dE\,E\,n(E) = \int\!d^3r\!\int\!d^3v\,
    [\tfrac12 v^2+\phi(r)]\,f(\tfrac12 v^2+\phi(r)) \\ 
    &= {\cal K} + 4\pi \int_0^{r_0} \!dr\,r^2\rho(r)\phi(r) = {\cal K} + 2\,{\cal U} + \frac{GM^2}{r_0}\;.
  \end{split}
\end{equation*}
Hence
\begin{equation}\label{eq:E2E}
  {\cal E} = \tfrac13 M\Big(\bar E - \frac{GM}{r_0} \Big)\;.
\end{equation}
In our dimensionless setup we have
\begin{equation}\label{eq:sigma1}
  \sigma^2(r_\ast x) = |\phi_0|\,{\tilde\sigma}^2(u(x)) \;,\quad {\tilde\sigma}^2(u) =
  \frac1{\nu(u)}\int_u^1\!ds\, F(s)\,(s-u)^{3/2}
\end{equation}
and
\begin{equation*}
  {\cal E} = \tfrac12 M \phi_0\, \eta\;,
\end{equation*}
where
\begin{equation}\label{eq:eta1}
  \eta = \int_0^1\!du\,x^3(u)\nu(u) = 2\!\int_0^1\!du\,x'(u)x^2(u)\nu(u){\tilde\sigma}^2(u) 
\end{equation}
is the dimensionless average of the squared velocity dispersion and plays the role of a
dimensionless temperature.

Eq.~\eqref{eq:E2E} now takes the form
\begin{equation}\label{eq:eta2}
    \eta  = \tfrac23 m\Big( 1-\bar s + \frac{m}{x(1)} \Big) \;,
\end{equation}
where
\begin{equation}\label{eq:sbar}
  \bar{s} = 1 - \frac{\bar{E}}{\phi_0} = \frac1{m}\int_0^1\!ds\,s N(s)
\end{equation}
is given by the explicit expression eq.~\eqref{eq:bars} in the DARKexp model. On the other hand,
both $m$ and $x(1)$ are not known as functions of $b$ before a solution of
eq.~\eqref{eq:xixi} is found.

{\highlight
\section{Analytic results}\label{sec:anres}

As a matter of fact, we do not know how to analytically find the general solution of the DARKexp
problem, as defined by eqs.~\eqref{eq:chain} and~\eqref{eq:xixi}. Some analytical results were
obtained in \cite{destri2018} concerning the next--to--leading asymptotics near $s=0$ of $F(s)$,
$\nu(s)$, $G(s)$, and $x(s)$. In the present formulation, upon introducing functions $\xi_\nu(s)$
and $\xi_x(s)$ by
\begin{equation}\label{eq:nux2xi}
    \nu(s) = 3\,\xi_\nu(s)  \;,\quad x(s) = (2s)^{1/2}(1-h s)^{-1}\xi_x(s) \;,
\end{equation}
these results read 
\begin{equation}\label{eq:nnlasy} 
  \begin{split}
   \xi_F(s) &\simeq 1 - \tfrac{32}{45} K s^{1/2} -\tfrac2{45}K^2s\log s  +O(s) \;, \\[1ex]
   \xi_\nu(s) &\simeq 1 - \tfrac13\pi K s^{1/2} - \tfrac{16}{135}K^2s\log s  +O(s) \;, \\[1ex]
   \xi_x(s)  &\simeq 1 + \tfrac{1}{12}\pi K s^{1/2} + \tfrac4{225}K^2s\log s +O(s) \;, \\[1ex]
   \xi_G(s) &\simeq 1 + \tfrac{32}{45} K s^{1/2} +\tfrac2{45}K^2s\log s +O(s)\;.
  \end{split}
\end{equation} 
Moreover, in \cite{destri2018} it was also argued that the next--to--leading asymptotics
\eqref{eq:nnlasy} is just the beginning of a complete, possibly asymptotic expansion of the form
\begin{equation}\label{eq:asympt}
  \xi(s) = 1 + \sum_{n=0}^\infty\sum_{k=0}^n\big(a_{nk}+b_{nk}s^{1/2}\big)s^n\log^{n-k}\!s \;,
\end{equation}
where $a_{00}=0$ by construction. The other coefficients are unknown, except for $b_{00}$ and
$a_{10}$ which are given by eq.~\eqref{eq:nnlasy} in terms of the global normalization parameter
$K$. We will make use of eq.~\eqref{eq:asympt} in section \ref{sec:afit} to check the agreement
of our numerical results with eq.~\eqref{eq:nnlasy}.

Another (partial) analytic result is possible in the limit $b\to +\infty$, when by
eq.~\eqref{eq:bars} ${\bar s}\to1$ and the average one--particle energy ${\bar E}\to0$. If the mass
density stays regular in this limit, thus keeping the total mass finite, then $N(s)$ must
concentrate at $s=1$ [that is $n(E)$ at $E=0$] while $G(s)$ remains regular. Hence it is $F(s)$ that
concentrates at $s=1$ leading, by eq.~\eqref{eq:F2nu} and the requirement that $\xi_\nu(0)=1$, to the
explicit analytic result
\begin{equation}\label{eq:xinfty}
  \lim_{b\to+\infty} \xi_\nu(u) = (1-u)^{1/2}\;,
\end{equation}
which is compatible with the asymptotics \eqref{eq:nnlasy} because the normalization constant $K$
[see eq.~\eqref{eq:Km}] vanishes in the limit. In dimension--full variables we have
\begin{equation*}
  \lim_{b\to+\infty} \rho(r) = \rho_0[\phi(r)/\phi_0]^{1/2}\;,
\end{equation*}
so that Poisson's eq.~\eqref{eq:pois1} becomes the Lane--Emden equation with $n=1/2$, which does not
have a known analytic solution and describes a finite--size system \cite{bt}. But it is easy to
solve it numerically, for instance in our preferred form \eqref{eq:pois2} [more precisely in the
form \eqref{eq:pois3} introduced in section \ref{sec:intranf}]. The corresponding values of some
relevant quantities, such as the dimensionless mass $m$ and size $x(1)$, are listed in the lowest
row of Table \ref{table:table1}. Notice that in the limit $b\to +\infty$, eq.~\eqref{eq:FG} only
constrains the end value at $s=1$ of $G(s)$, namely $G(1)=3m$. This can be alternatively computed
directly from $x(u)$ through Gauss's law $mx'(1)= x^2(1)$. The agreement of the these two
determinations of $m$ provides a measure of the accuracy of our numerical solution of the DARKexp
problem and will be systematically checked for many finite values of $b$ in section \ref{sec:results}.  
}

\section{Numerical approach}\label{sec:numapp}

More insight on the DARKexp problem appears to be possible only through some numerical approach,
which we describe in this section. In ref.~\cite{williams2010a}, within a different setup and for
the distinct case of $1/r-$cusped density profiles, rather curious numerical results were reported,
but without details on the specific method employed.

\smallskip
We first observe that eqs.~\eqref{eq:nnlasy} suggest a change of variable from $s$ to $z=s^{1/2}$,
so that $\xi(z^2)$, for $\xi=\xi_F,\,\xi_\nu,\,\xi_x,\,\xi_G$, becomes differentiable in $z=0$. As a
matter of fact, one can consider also higher powers of $z$ in the reparametrization, such as $s=z^p$
with $p>2$, making $\xi(z^p)$ differentiable even further in $z=0$. In our actual computer programs
we chose $p=4$, namely $s=z^4$.

Smoothness near $s=0$ plays an important role for numerical accuracy. For instance, even if the
$s^{-1/2}$ singularity, which appears in the integrand of eq.~\eqref{eq:F2nu} when $u=0$, is
integrable, it is the right neighborhood of $s=0$ that gives the most important contribution to
$\nu(u)$ when $u\gtrsim0$ and the reparametrization $s=z^p$ improves accuracy in the
integration. Furthermore, by tuning the value of the free parameter $h$, we can make each $\xi(z^p)$ 
to have a variation of order one on the interval $[0,1]$, besides being certainly smooth there. 
It follows that $\xi(z^p)$ can be approximated very accurately and effectively by an expansion 
in a relatively small number of scaled and shifted Chebyshev polynomials $T_n(2z-1)$.

\smallskip 
Hence our numerical approach starts by discretizing the interval $[0,1]$ using $N+1$
Chebyshev points of the second kind, according to
\begin{equation}\label{eq:chegrid}
  s_k=z_k^p \;,\quad z_k = 2\,\sin^2\!\Big(\frac{k\pi}{2N}\Big) \;,\quad k=0,1,\ldots,N \;, 
\end{equation}
so that the function $\xi(s)$ is replaced by the discrete array $\{\xi(s_k),\,k=0,1,\ldots,N\}$. At
the same time, this array uniquely defines, for all $s\in[0,1]$, a polynomial approximation to the
exact $\xi(s)$ as
\begin{equation}\label{eq:chebexp}
  \xi(z^p) \approx \sum_{n=0}^N c_n T_n(2z-1) \;,
\end{equation}
which allows an accurate calculation of the integral transforms in eqs.~\eqref{eq:F2nu} and
\eqref{eq:x2G}. These transforms read for the $\xi$ functions
\begin{equation}\label{eq:xiF2xinu}
  \xi_\nu(u) = \tfrac13 K\int_u^1\!ds\,s^{-1}(s-u)^{1/2}{\highlight (1-h s)}^{5/2}\,\xi_F(s) \;.
\end{equation}
and
\begin{equation}\label{eq:xix2xiG}
  \xi_G(s) = \frac8{3\pi s^2}{\highlight (1-h s)}^{5/2}\int_0^s\!du\,(s-u)^{-1/2}
  \bigg[\frac{u^{1/2}\xi_x(u)}{\highlight 1-h u}\bigg]^3\;,
\end{equation}
The map from the function values $\xi(s_k)$ [or local functions of them as in the
r.h.s. of eq.~\eqref{eq:xix2xiG}] to the Chebyshev coefficients $c_n$ can be performed very
efficiently through Clenshaw--Curtis formulas (\emph{a.k.a.} fast discrete cosine transform).

\smallskip
Let us also observe that, by construction, $\xi(0)=1$ for all four
$\xi_F,\,\xi_\nu,\,\xi_x$ and $\xi_G$, so that there are only $N$ independent unknowns, that we
choose to be those of the array $\{\xi_F(s_k),\,k=1,\ldots,N\}$.  

\smallskip

Quite obviously, the accuracy of the numerical approach described in the rest of this section
depends on $N$, that is on the size of the discretization grid over the interval of interest
$[0,1]$.  Since the execution speed on any computer rapidly increases for increasing $N$, some trade
off is required. Thanks to the excellent properties of Chebyshev interpolation, it turns out that a
very good accuracy can be attained, for a whole range of $b$ values, already with $N=400$, while
keeping the required computational resources well within those of a powerful workstation. We have
estimated this accuracy to be better than $10^{-9}$ for all grid values, by comparison with
the results obtained when $N=10^3$ and by other means described in the Appendix.

\subsection{Integral transforms and Poisson's equation}\label{sec:intranf}

The calculation of the integral transforms in eqs.~\eqref{eq:xiF2xinu} and \eqref{eq:xix2xiG} is
made fast, fully preserving accuracy, by pre--computing the relevant integrals with the Chebyshev
polynomials $T_n(2z-1)$. For instance, for eq.~\eqref{eq:xiF2xinu} we have
\begin{equation*}
    \xi_\nu(z^p) \approx K \sum_{n=0}^N c_{F,n}\, I^{(F\nu)}_n(z)\;, 
\end{equation*}
where
\begin{equation}\label{eq:IFnu}
  I^{(F\nu)}_n(z) = \tfrac13 p \int_z^1\!dy\,y^{-1}(y^p-z^p)^{1/2}{\highlight (1-h z^p)}^{5/2}\,T_n(2y-1) \;.
\end{equation}
Hence we may set
\begin{equation}\label{eq:IFnu1}
  \xi_\nu(s_k) \approx \sum_{n=0}^N c_{\nu,n} T_n(2z_k-1) = K \sum_{n=0}^N c_{F,n}\,I^{(F\nu)}_{nk}
  \;, \quad I^{(F\nu)}_{nk}\equiv I^{(F\nu)}_n(z_k)\;,
\end{equation}
from which, if necessary, the $c_{\nu,n}$ can be recovered by the fast Clenshaw--Curtis
algorithm. The important point is that only a vector--matrix multiplication is needed in this
$\xi_F\longrightarrow \xi_\nu$ step, since the matrix entries $I^{(F\nu)}_{nk}$ can be computed, for
a given choice of $N$, once and for all. It should also be noticed that the constant $K$ is
fixed by the requirement that $\xi_\nu(0)=1$, that is
\begin{equation*}
  K^{-1} = \tfrac13 \int_0^1\!ds\,\frac{{\highlight (1-h s)}^{5/2}}{s^{1/2}}\,\xi_F(s)
  \approx \sum_{n=0}^N c_{F,n}\,I^{(F\nu)}_{n0} \;.
\end{equation*}

The second step $\nu\longrightarrow x$ in the chain \eqref{eq:chain} requires to numerically solve
Poisson's equation, in either one of the two forms in eq.~\eqref{eq:pois1} or
eq.~\eqref{eq:pois2}. In our setup the latter form is more appropriate, since the grid is fixed in
the $u$ variable. Using the form of eq.~\eqref{eq:pois1} where $x$ is the independent variable has
drawbacks for accuracy and/or speed, especially if $x(1)$ happens to be very large.  On the other
hand, in the form of eq.~\eqref{eq:pois2} where $u$ is the independent variable, high accuracy is
possible also when $x(1)$ becomes very large, but special care in needed for the initial conditions
$x \simeq (2u)^{1/2}$ as $u\to0$.  We found it convenient to reformulate eq.~\eqref{eq:pois2} as
follows
\begin{equation}\label{eq:pois3}
  \xi_x(e^y) = {\highlight (1-h\,e^y)}e^{\gamma(y)/2} \;,\quad
  \gamma'' = \tfrac32 + 2\gamma'+\tfrac12\gamma'^2-\tfrac32\xi_\nu(1+\gamma')^3e^{\gamma} \;.
\end{equation}
The initial conditions on $\gamma$ now read $\lim_{y\to-\infty}(\gamma,\gamma')=(0,0)$ and can be
implemented numerically by setting $(\gamma,\gamma')=(0,0)$ at any large negative value $y_0$ such
that $e^{y_0}$ vanishes as double--precision floating number. Notice also that values of
$\xi_\nu(u)$ at generic values of $u$ are needed to solve eq.~\eqref{eq:pois3} and the most
effective interpolation is just the Chebyshev one.

The final outcome is the array $\{\xi_x(s_k),\,k=0,1,\ldots,N\}$, which is used in the last step
$x\longrightarrow G$ with the same reduction of the integral transform \eqref{eq:xix2xiG} to a
matrix--vector multiplication as in the first step. The only caveat, from the numerical point of
view, is to avoid the multiplication of the possibly very large/small prefactor $s^{-2}(1+r
s)^{5/2}$ by the possibly very small/large integral when $s\to0$ or both $s$ and $h$ are close to
$1$. This is accomplished by the following change of integration variable
\begin{equation*}
  u = u(t,s,h) = s(1-t^2)\frac{1+h(1-s)}{1+h(1-s-t^2)} \;,
\end{equation*}
which transforms eq.~\eqref{eq:xix2xiG} into the smoother expression
\begin{equation}\label{eq:smooth}
  \xi_G(s) =  \frac8{3\pi}[1+h(1-s)]^{5/2}\int_0^1\!dt \bigg[
  \frac{(1-t^2)^{1/2}}{1+h(1-t^2)(1-s)}\,\xi_x\big(u(t,s,h)\big) \bigg]^3 \;.
\end{equation} 
Then we can write
\begin{equation*}
    [\xi_x(z^p)]^3 \approx \sum_{n=0}^Nc_{x,n} T_n(2z-1)\;,\quad 
    \xi_G(s_k) \approx \sum_{n=0}^N c_{x,n}\,I^{(xG)}_{nk} \;,
\end{equation*}
where the pre--computable matrix entries read
\begin{equation}\label{eq:IxG} 
    I^{(xG)}_{nk} =  \frac{8}{3\pi} [1+h(1-s_k)]^{5/2}\int_0^1\!dt \bigg[
  \frac{(1-t^2)^{1/2}}{1+h(1-t^2)(1-s_k)}\,T_n\big(2[u(t,s_k,h)]^{1/4}-1\big) \bigg]^3 \;.
\end{equation}
As discussed in the Appendix, the accuracy of this approach crucially depends on the convergence
toward zero of the coefficients $c_{x,n}$ as $n$ grows. In turns, this depends on the behavior of
$\xi_x(s)$ as $s\to1$. The behavior as $s\to0$ is not as important, in spite of the logarithmic
singularity in eq.~\eqref{eq:nnlasy}, because of the extra powers of $u$ appearing in the integrand
of eq.~\eqref{eq:xix2xiG}. We see that the free parameter $h$ can be fixed beforehand to guarantee that
$\xi_x(1)$ does not becomes too large, thus preventing the risk of a poor convergence of the coefficients
$c_{x,n}$.

\smallskip {\highlight A worthwhile byproduct of the smoother rewriting \eqref{eq:smooth} of the
  map from $\xi_x$ to $\xi_G$ is the rigorous derivation of the large--distance behavior, stated
  without details in section \ref{sec:infinite}, of the mass density in a system with infinite
  size. In such a system, which by assumption has a finite total mass $M$, the gravitational
  potential $\phi(r)$ dies off at radial infinity as $M(4\pi r)^{-1}$ (recall that $4\pi G=1$). In
  the dimensionless setup this reads $x(u)\simeq m(1-u)^{-1}$ as $u\to1$. Then, setting $h=1$, from
  eq.~\eqref{eq:nux2xi} we see that $\xi_x(1)=2^{-1/2}m$ is finite and that a cubic pole as $u\to1$
  appears in the integrand of eq.~\eqref{eq:xix2xiG} when $s\to1$. The smoother rewriting
  \eqref{eq:smooth} explicitly shows that this exactly compensate the the zero coming from the
  prefactor $(1-s)^{5/2}$ in eq.~\eqref{eq:xix2xiG} when $h=1$. Hence $\xi_G(1)$ is finite and by the
  fundamental equation \eqref{eq:xixi} we see that $\xi_F(1)$ must be finite and larger than
  zero. Then a simple power counting in eq.~\eqref{eq:xiF2xinu} shows that $\xi_\nu(u)\sim (1-u)^4$
  as $u\to1$ or, equivalently, $\rho(r)\sim r^{-4}$ as $r\to\infty$. }

\medskip
Altogether, the method just described turns out to be fast and accurate in computing the numbers
$\{\xi_G(s_k),\,k=0,1,\ldots,N\}$ from given numbers $\{\xi_F(s_k),\,k=0,1,\ldots,N\}$, that is
recovering the approximated $\xi_G(s)$ from a given approximated $\xi_F(s)$. In principle, one could
consider also the reverse computation along the chain eq.~\eqref{eq:chain}, that is from $\xi_G(s)$
to $\xi_F(s)$, with the apparent advantage of reducing eq.~\eqref{eq:pois2} to an explicit
calculation of $\nu$ from $x$. There are two shortcomings of this reverse approach: differentiation
introduces larger numerical errors than integration and, most importantly, there is no guarantee of
preserving positivity when going from an arbitrary positive $\xi_G(s)$ back to $\xi_F(s)$.

\subsection{System of non--linear equations}\label{sec:nlse}

With $\xi_G$ accurately computable from $\xi_F$, eq.~\eqref{eq:xixi} becomes,
in our discretized setup, a system of $N$ non--linear equations for the $N$ unknowns
$\xi_F(s_k)$. We need to numerically solve these equations, which means to make the (relative)
residuals
\begin{equation}\label{eq:residuals}
  R_k[\xi_F] = \frac{\xi_F(s_k)\xi_G(s_k)}{A(bs_k)}-1 \;,\quad k=1,2,\ldots,N\;,
\end{equation}
as small as possible. On general grounds, due to numerical roundoffs, the magnitudes $|R_k|$ cannot
be smaller than the machine epsilon, namely $2.22044...\cdot10^{-16}$ in double--precison. But any
algorithm that tries to minimize the $|R_k|$, assuming that there exist an exact, analytic solution,
will stop at values that can be much larger than that, depending on $N$, on the $N-$dimensional
numerical neighborhood of the solution, on the algorithm itself and on how sensible it is on the
accuracy in the calculation of the chain eq.~\eqref{eq:chain}. The problem is to determine when
small is small enough, since we should expect that, by varying $b$, certain exact solutions of
eq.~\eqref{eq:xixi} might appear, disappear, coalesce or more simply become more difficult to
pinpoint with numbers. In principle, one would like to disentangle, if necessary, the case when the
numerical $\xi_F$ is a poor approximation of an exact solution from the case when there no exact
solution at all. This is a difficult task and, in the absence of analytic results concerning the
existence of solutions for any given value of $b$, it is a numerical task beyond the scope of this
work. Here we want to present a class of convincing numerical solutions, rather than rule out the
existence of solutions for certain values of $b$. In other words, we will not state that there exist
no solution when $b$ takes values for which we cannot not produce accurate numerical solutions.

An accurate numerical solution, as we define it, is any output of a program devised to minimize the
residuals that succeeds to the point that
\begin{equation}\label{eq:success}
  {\cal R} \equiv \underset{k}{\rm max}\,|R_k| < \epsilon \;,  
  \quad \epsilon \equiv 10^{-11} \;.  
\end{equation}
The specific value of $\epsilon$ was empirically determined as the typical one worth attaining and
attainable upon optimization of all steps of the calculation.  In other words, we found that
$\epsilon=10^{-11}$ is close enough to the limit of the double--precision accuracy for the problem at
hand. More details on this result can be found in the next section and in the Appendix.
However, while very often we obtained residuals much smaller than eq.~\eqref{eq:success}, with
${\cal R}$ even of order $10^{-14}$, we regard as acceptable solutions also those obtained in a
special class of cases in which the bound \eqref{eq:success} is not satisfied, although the
residuals are as small as $10^{-8}$. The reason for this exception to eq.~\eqref{eq:success} is that
we can identify quite precisely the source of extra numerical errors causing the degraded accuracy
and that these \emph{reasonably inaccurate} numerical solutions fit very well into a continuous
$b-$dependent family.

Let us now describe the two numerical approaches we use to deal with eq.~\eqref{eq:xixi}.

The first, straightforward approach is to start from some initial $\xi_F^{(0)}$ and
then iterate as follows
\begin{equation}\label{eq:simplemap}   
  \xi_F^{(n+1)}(s_k) = (1-\alpha) \xi_F^{(n)}(s_k) + \alpha\frac{A_b(s_k)}{\xi_G^{(n)}(s_k)}
  \;,\quad k=1,2,\ldots,N \;,
\end{equation}
with $\alpha$ a damping parameter that might help convergence to some fixed point, which would then
be a solution of our problem. Of course, this method can work only if there are stable fixed points,
that is such that the Jacobian matrix 
\begin{equation}\label{eq:jac}
  J_{kn}(\alpha) =  (1-\alpha)\delta_{kn} + \alpha A_b(s_k)\frac{\partial [\xi_G(s_k)]^{-1}} 
  {\partial \xi_F(s_n)}  \;,
\end{equation}
when evaluated at such fixed point, has eigenvalues with magnitudes all less than one for some value
of $\alpha$. Even so, the iteration will converge only if the initial $\xi_F^{(0)}$ is in the
basin of attraction of some stable fixed point. Moreover, due to numerical roundoffs, there cannot
be true convergence to a fixed point. Rather, the sequence $\xi_F^{(n)}$ will indefinitely and more
or less randomly and tightly fluctuate around the stable fixed point profile, provided the residuals
are small enough. It will become apparent in the next section that residuals satisfying
eq.~\eqref{eq:success} are indeed small enough.

\smallskip
The simple iteration method should be tried first. If it fails we can resort to the second
approach, possibly much slower but safer and much richer of tunable parameters, that is some
general--purpose algorithm for systems of non--linear equations (see \emph{e.g.}~\cite{nocedal}). For
instance one that, starting from an initial $\xi_F^{(0)}$, iteratively minimizes some
positive--definite scalar expression that vanishes at the solution, such as
\begin{equation}\label{eq:f2min} 
  {\cal F}[\xi_F] = \sum_{k=0}^N w_k(R_k[\xi_F])^2 \;, \quad w_k > 0 \;,
\end{equation}
or any other scalar, positive--definite expression that vanishes when $R_k=0$. This
quantity is usually called the \emph{objective function}.

The obvious problem of any such algorithm is that it can only find local minima of the objective
function, which could in principle be different for different starting points. The actual solutions
of eq.~\eqref{eq:xixi} for any given $b$ correspond to global minima where the residuals vanish in
the numerical sense, namely that, according to our understanding, satisfy the criterion
\eqref{eq:success}.

The two approaches just described can be compared, when both are successful, to verify their
consistency if the starting $\xi_F^{(0)}$ is the same, since they have a different structure of
basins of attraction. They can also be mixed, to improve the overall speed and/or accuracy of the
calculation and to try and disentangle global minima from local minima of the objective function.

\smallskip 
Finally, let us consider the choice of the array $\xi_F^{(0)}$ on which to start the minimization
run. A convenient strategy is to find, by trial and error for at least one value $b=b_0$, an
acceptable solution. Next, this $\xi_F$ can be used as initial configuration for another run with a
new value $b=b_1$, close enough to $b_0$, relying on the continuity in $b$ of the solution. Then
this process can be repeated as many times as possible, perhaps adjusting each time the increment of
$b$ and/or using initial configurations obtained by extrapolation in $b$.  For the first run at
$b=b_0$ one could start for instance from the simplest initial array $\xi_F^{(0)}=1$, but any other
choice is possible, as long as it does lead to an acceptable solution. The results reported in
section \ref{sec:results} have been obtained using this strategy with $b_0=2$ and $\xi_F^{(0)}=1$.

\section{Numerical results}\label{sec:results}

\begin{figure}[tbp]
  \begin{center}
    \includegraphics[width=.68\textwidth]{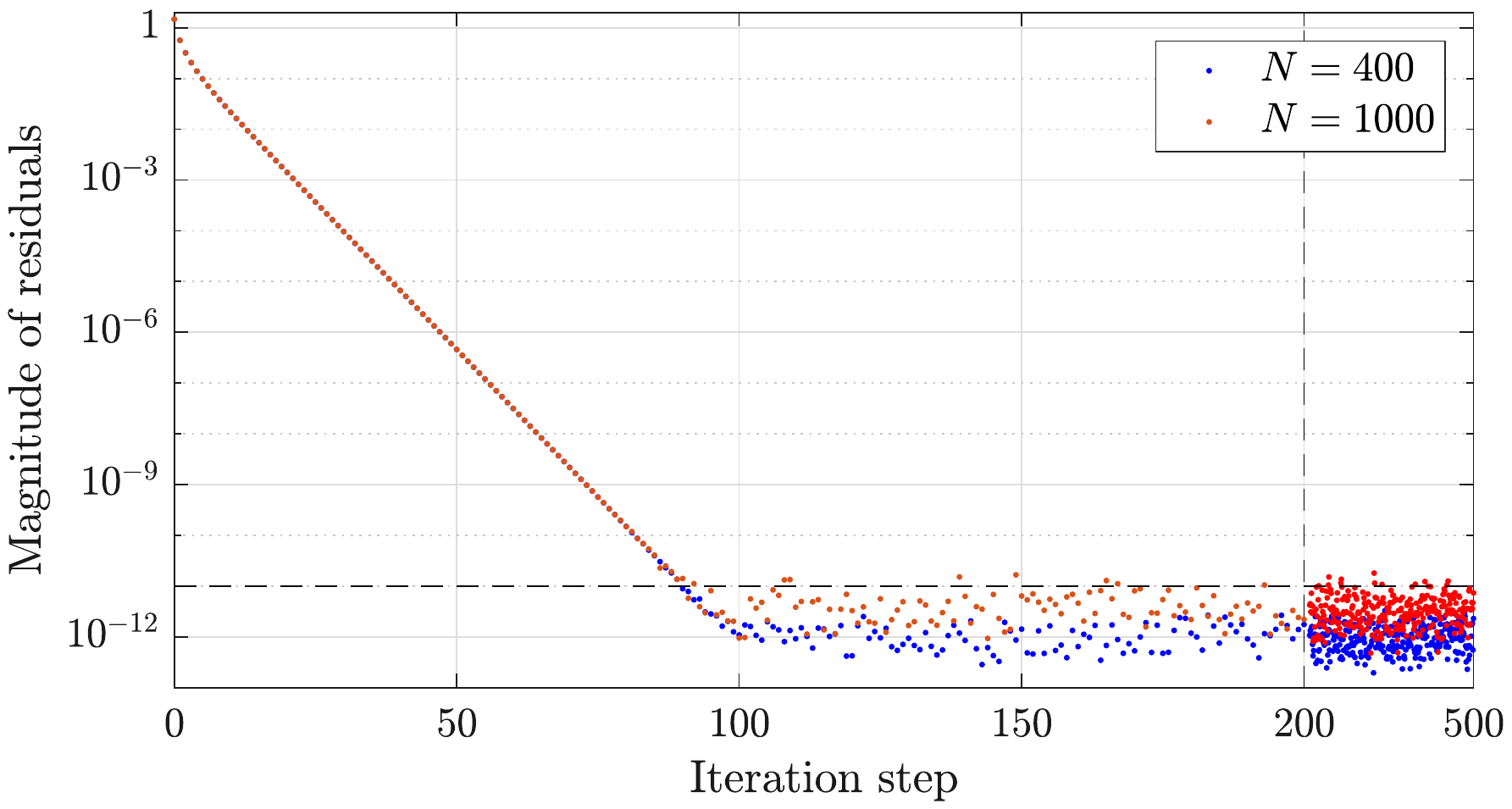}
    \caption{The value of ${\cal R}$ [see eqs.~\eqref{eq:residuals} and~\eqref{eq:success}] at each
      iteration of the map \eqref{eq:simplemap}. Almost all values in the fluctuating plateau satisfy 
      the criterion \eqref{eq:success}.}
    \label{fig:simplemap}
  \end{center}
\end{figure}

\begin{figure}[tbp]
  \begin{center}
    \includegraphics[width=.8\textwidth]{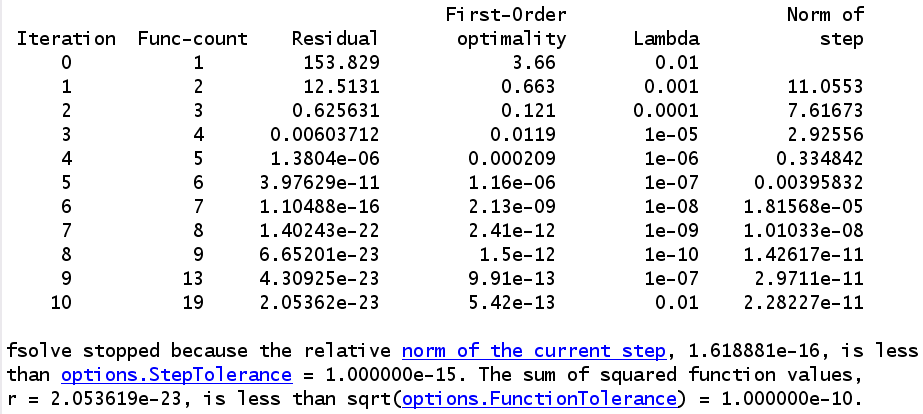}
    \caption{Output of MATLAB {\tt fsolve} running the Levenberg--Marquardt algorithm when $b=2$ and
      $N=400$. The 'Residual' column exhibits the values of the objective function ${\cal F}[\xi_F]$
      in eq.~\eqref{eq:f2min} with unit weights $w_k$. Notice the little irrelevant MATLAB bug in the
      output, namely the final 'Residual' compared to the square root of {\it FunctionTolerance}
      rather than {\it FunctionTolerance} itself.}
    \label{fig:iter_val1}
  \end{center}
\end{figure}

\begin{figure}[tbp]
  \begin{center}
    \includegraphics[width=.68\textwidth]{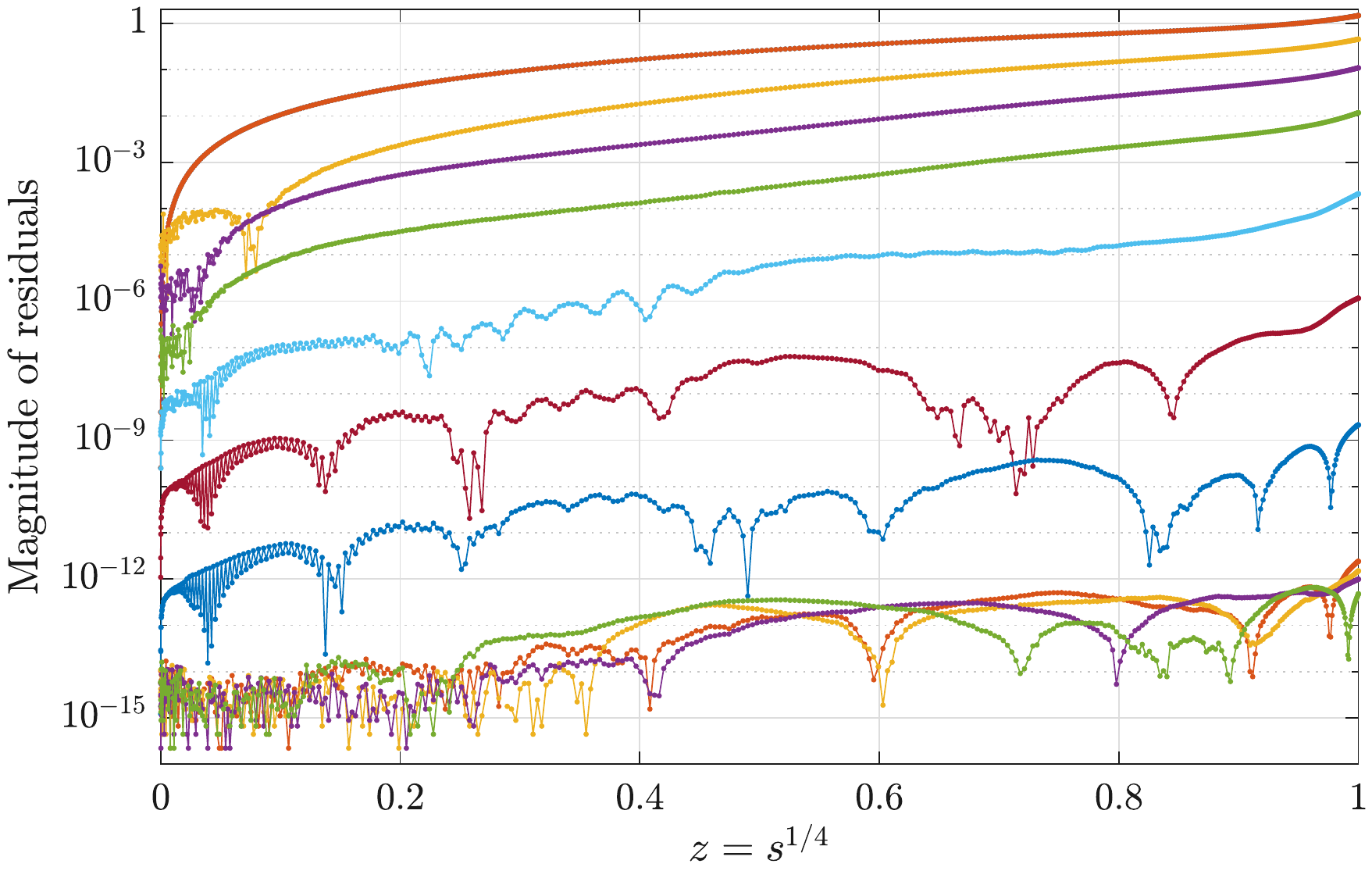}
    \caption{The absolute values of the residuals $R_k$ [see eq.~\eqref{eq:residuals}] at each
      iteration of {\tt fsolve} running as in Fig~\ref{fig:iter_val1}. This is a successful run
      according to the criterion \eqref{eq:success}. Notice the appearance of rugged
      fluctuations which are typical of numerical roundoffs.}
    \label{fig:iter_resid1}
  \end{center}
\end{figure}

We begin by describing in detail, when $b=2$, the computation of a profile $\xi_F$ satisfying
eq.~\eqref{eq:xixi}, in the sense of criterion \eqref{eq:success}, using both methods described in
the previous section. The initial profile is $\xi_F^{(0)}(s)=1$ and the free parameter $h$ is
momentarily set to $0$.  In this case the iterative map \eqref{eq:simplemap} converges rather rapidly
already with $\alpha=1$. Just as successful, albeit slower, is the minimization of the objective
function ${\cal F}[\xi_F]$ in eq.~\eqref{eq:f2min} with the weights $w_k\equiv 1$.

\noindent {\bf First method}\\
Using $N=400$ Chebyshev points, after less than $100$ iterations of the map \eqref{eq:simplemap},
the monotonic decreasing trend of the residuals $R_k$ [see eq.~\eqref{eq:residuals}] has come to an
end. Then the $R_k$ keep fluctuating with values of order $10^{-12}$ [see
Fig.~\ref{fig:simplemap}]. We can take as solution any profile  $\xi_F$ satisfying
eq.~\eqref{eq:success}. No accuracy gain is obtained using $10^3$ Chebyshev points [see
Fig.~\ref{fig:simplemap}]. On the contrary, as should be expected if the calculation is close to its
double--precision limit, the fluctuating residuals with $N=10^3$ are slightly larger that those with
$N=400$, because of the increased number of roundoffs.

\smallskip
\noindent {\bf Second method}\\
The minimization algorithm we adopted is {\tt fsolve} in MATLAB R2017b. We first run it with $N=400$
points, until it stops because the proposed change in $\xi_F$ is too small [see
Fig.~\ref{fig:iter_val1} and~\ref{fig:iter_resid1}]. This is a successful run, since the final
residuals satisfy the condition \eqref{eq:success}. We again repeat the calculation with $10^3$,
with no increase of accuracy on the residuals.

\begin{figure}[tbp]
  \begin{center}
    \includegraphics[width=.68\textwidth]{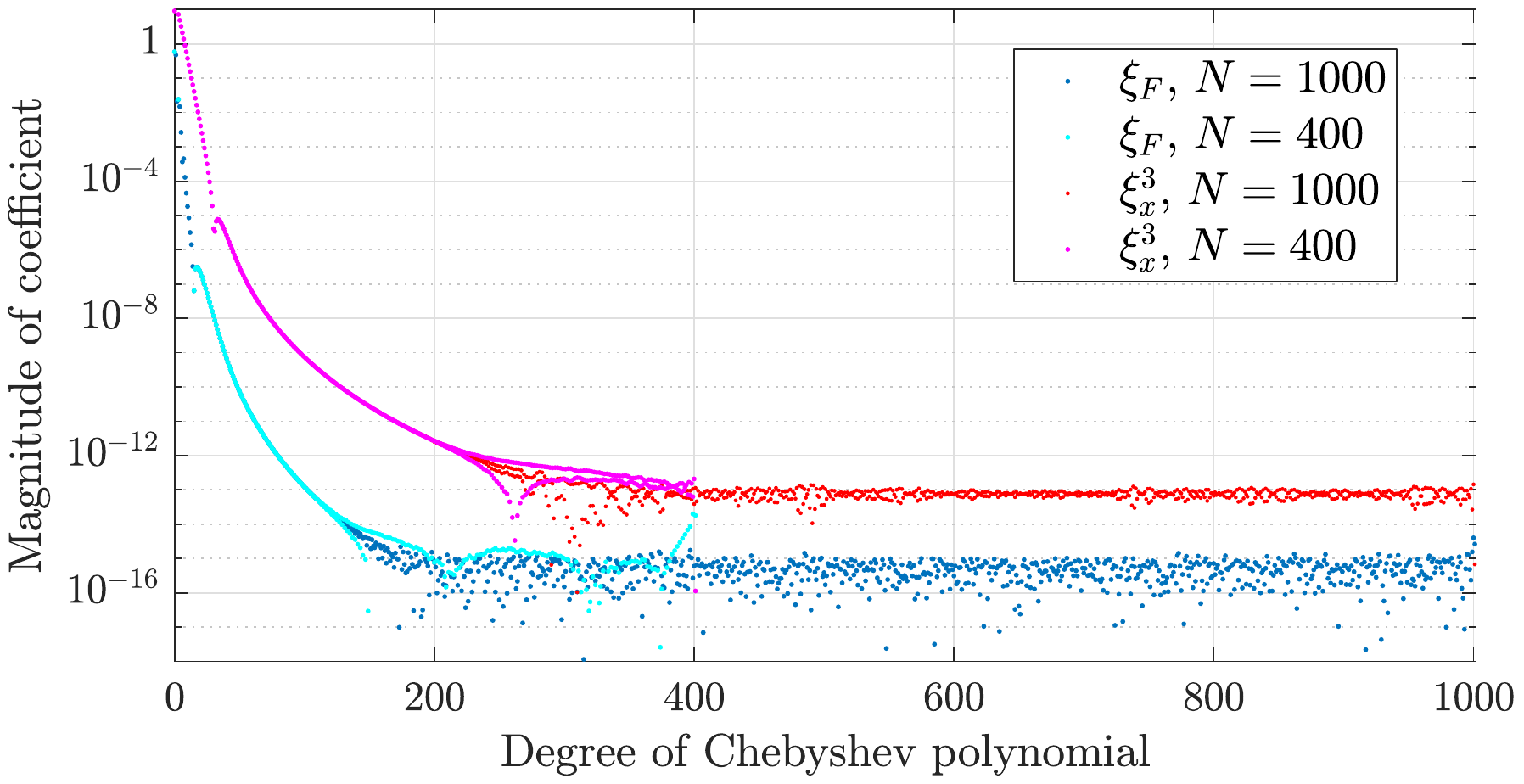}
    \caption{The Chebyshev coefficients of $\xi_F(z^4)$ and $\xi_x^3(z^4)$ when $b=2$. }
    \label{fig:coeffs2}
  \end{center}
\end{figure}

\medskip
In terms of the Chebyshev interpolating polynomials, we find that 
\begin{equation*}
  \underset{s}{\rm max}\big|\xi_F^{(N=400)}(s) - \xi_F^{(N=1000)}(s)\big| \sim  10^{-11}\;,
\end{equation*}
for both methods. The excellent agreement between the two calculations, with $N=400$ and $N=10^3$,
can be traced back to the fast decay of the Chebyshev coefficients of $\xi_F(z^4)$ and
$\xi_x^3(z^4)$, which are plotted in Fig.~\ref{fig:coeffs2}. Denoting with $\xi_F^{(A)}$ and
$\xi_F^{(B)}$ the profiles obtained by the two methods, we find that
\begin{equation*}
  \underset{k}{\rm max}\big|\xi_F^{(A)}(s_k) - \xi_F^{(B)}(s_k)\big| \sim 10^{-11}\;,
\end{equation*}
both for $N=400$ and $N=1000$. Another check on the stability of the calculation is possible by
varying the free parameter $h$. We verify that $(1+h s)^{5/2}\xi_F(s,h)$ indeed does not depend on
$h$, within order $10^{-13}$, as long as $h$ is not too close to $1$. When $h$ gets too close to
$1$, $\xi_F(1,h)$ becomes very large, compromising the accuracy of the Chebyshev
approximation. Altogether, according to the analysis of the Appendix, we expect our numerical
$\xi_F$ to differ at most by order $10^{-10}$ from the exact solution.

\begin{figure}[tbp]
  \begin{center}
    \includegraphics[width=.68\textwidth]{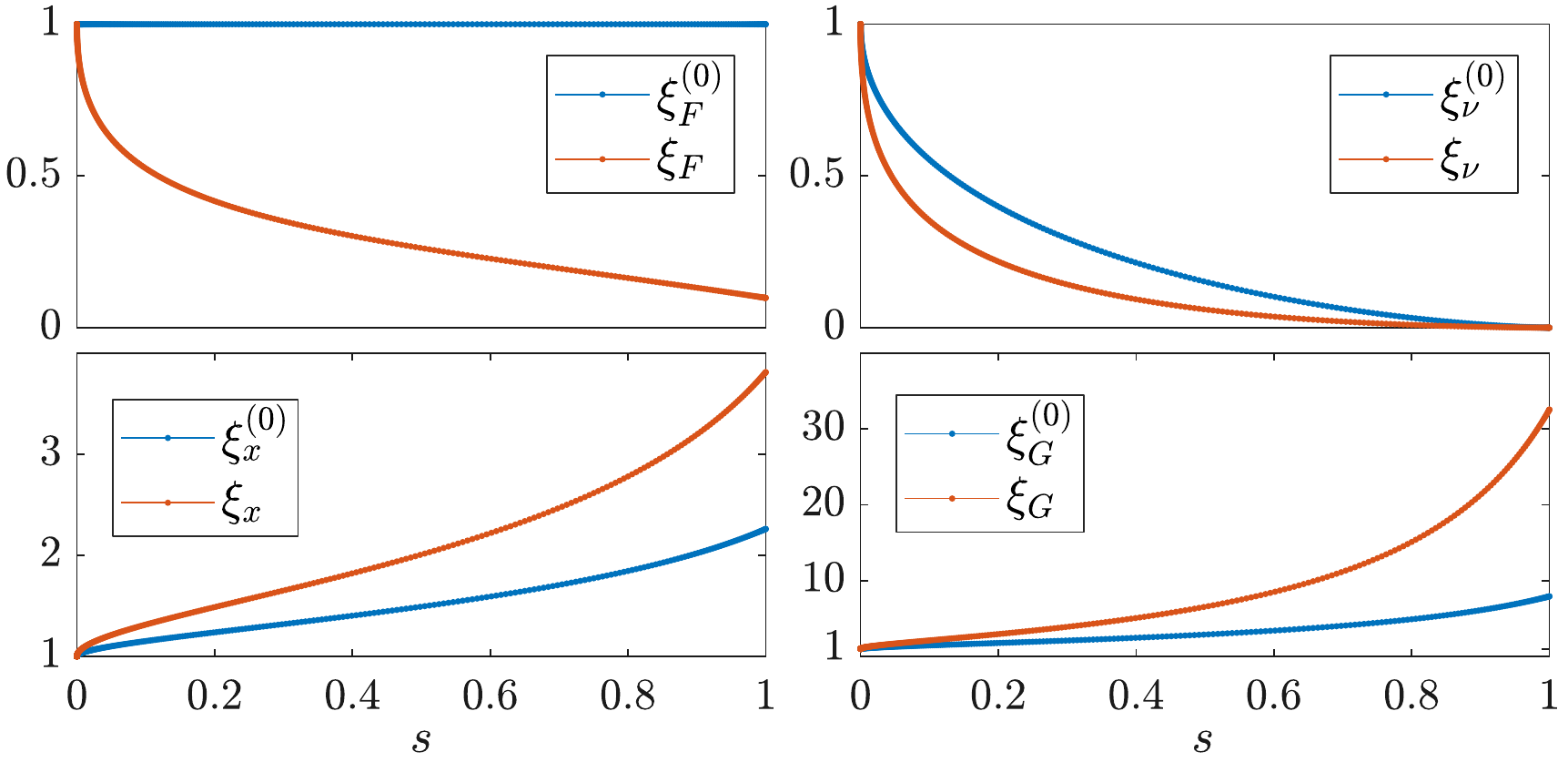}
    \caption{Initial and final profiles when $b=2$.}
    \label{fig:fourplots}
  \end{center}
\end{figure}

The initial and final profiles of $\xi_F$, $\xi_\nu$, $\xi_x$ and $\xi_G$, as functions of $s$ are
plotted in Fig.~\ref{fig:fourplots}.

\subsection{Solutions of the first kind}\label{sec:firstkind}

\begin{figure}[tbp]
  \begin{center}
    \includegraphics[width=.75\textwidth]{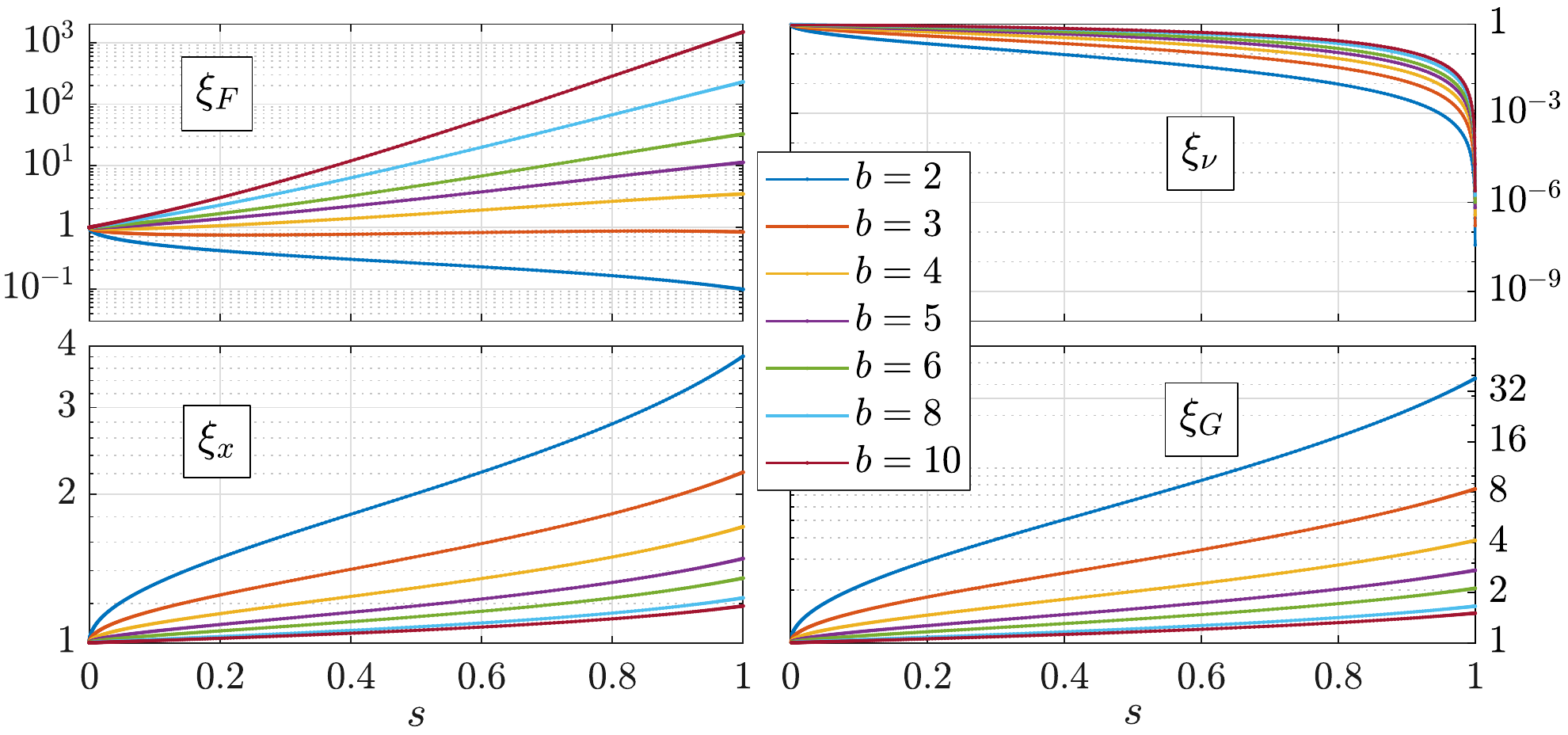}
    \caption{The four profiles $\xi_F$, $\xi_\nu$, $\xi_x$ and $\xi_G$ at selected values of $b\ge2$,
      with $h=0$ [see eqs.~\eqref{eq:F2xi}, \eqref{eq:G2xi} and \eqref{eq:nux2xi}. These are
      profiles corresponding to solutions of the first kind.}
    \label{fig:profiles1}
  \end{center}
\end{figure}

To compute solutions for other values of $b$ we follow the strategy outlined at the end of the
previous section. We use the final $\xi_F$ with $b=2$ as input for a new run with, say, $b=2.1$,
using either one of the two methods. If this run is successful, we have a solution at the new value
$b=2.1$. If not, we reduce the increment of $b$ until a successful run is realized. It turns out
that increasing $b$ from the initial value $b=2$ is quite easy and the incremental step can be much
larger than $0.1$. Even more, we find that one can always start from the initial profile
$\xi_F^{(0)}(s)=1$ with only a little extra cost in computer time. This is due to the behavior of
$\xi_F,\,\xi_\nu,\,\xi_x$ and $\xi_G$ as functions of $b$, as evident from
Fig~\ref{fig:profiles1}. The size $x(1)$ shrinks with increasing $b$, further improving accuracy and
convergence rate to the solution.

\smallskip
In Table \ref{table:table1} we list the number of iterations $N_\mathrm{iter}$, the value of the
residual $R$ and the values of the  of $K$, $x(1)$, $m$ and the dimensionless temperature
$\eta$ [see eq.~\eqref{eq:eta1}] for some selected values in the interval $2\le b\le 100$. Notice that
$m$ can be obtained in two, numerically inequivalent ways:
\begin{description}
  \item{$m_1\;$:} in terms of $K$ and $b$ as in eq.~\eqref{eq:Km}, with an accuracy that depends on
    how small are the residuals and how are they arranged around $0$;
  \item{$m_2\;$:} in terms of $x(1)$ and $x'(1)$, relying on Gauss's law, $mx'(1)= x^2(1)$, with an
    accuracy directly related to that of the solution of Poisson's equation.
\end{description}
A similar observation applies to $\eta$:
\begin{description}
  \item{$\eta_1\;$:} in terms of $m_1$, $\bar s$ and $x(1)$, as in eq.~\eqref{eq:eta2};
  \item{$\eta_2\;$:} from the integral in eq.~\eqref{eq:eta1}, that is $\eta =
    12\cdot2^{3/2} \int_0^1\!dz\,z^9\xi_x^3(z^4)\xi_\nu(z^4)$, computed through Chebyshev
    quadrature.
\end{description}
The values listed in Table \ref{table:table1} are the mean values $m=\tfrac12(m_1+m_2)$ and
$\eta=\tfrac12(\eta_1+\eta_2)$, while $\Delta m=|m_1-m_2|$ and $\Delta\eta=|\eta_1-\eta_2|$. The
agreement between the two determinations of $m$ and $\eta$ is better than our accuracy estimate,
except when $b=\infty$. {\highlight The slightly worse accuracy in this case can be traced back to
  the square--root behavior of $\xi_\nu(u)$ near $u=1$ [see eq.~\eqref{eq:xinfty}], which makes
  Chebyshev quadratures less efficient, especially concerning the calculation of $\eta_2$.}

% We do not expect any significant change of behavior for $b>100$. Systems will just be more compact,
% more massive and hotter.

%\FloatBarrier

\begin{table}
\centering
\begin{tabular}{|r|c|c|c|c|c|c|c|c|}
\hline
{\small $b$} & {\small $N_\mathrm{iter}$} & {\small $\cal R$} & {\small $K$} & {\small $x(1)$} & {\small $m$} & {\small $\eta$} & {\small $\Delta m$} & {\small $\Delta\eta$} \\
\hline
{\small 2} & {\small 91} & {\small 6.60$\,\cdot10^{-13}$} & {\small 3.47182} & {\small 5.38834} & {\small 2.11565} & {\small 0.93765} & {\small 6.18$\,\cdot10^{-12}$} & {\small 3.39$\,\cdot10^{-12}$} \\
\hline
{\small 3} & {\small 34} & {\small 3.86$\,\cdot10^{-13}$} & {\small 1.83567} & {\small 3.13586} & {\small 1.82206} & {\small 0.99742} & {\small 4.08$\,\cdot10^{-12}$} & {\small 3.10$\,\cdot10^{-12}$} \\
\hline
{\small 4} & {\small 25} & {\small 3.10$\,\cdot10^{-13}$} & {\small 1.01445} & {\small 2.43178} & {\small 1.74643} & {\small 1.08028} & {\small 1.28$\,\cdot10^{-12}$} & {\small 3.52$\,\cdot10^{-12}$} \\
\hline
{\small 5} & {\small 19} & {\small 5.76$\,\cdot10^{-14}$} & {\small 0.54827} & {\small 2.09556} & {\small 1.73452} & {\small 1.16810} & {\small 1.65$\,\cdot10^{-13}$} & {\small 3.82$\,\cdot10^{-12}$} \\
\hline
{\small 7} & {\small 13} & {\small 2.93$\,\cdot10^{-14}$} & {\small 0.14406} & {\small 1.80740} & {\small 1.77751} & {\small 1.33088} & {\small 1.16$\,\cdot10^{-12}$} & {\small 4.76$\,\cdot10^{-12}$} \\
\hline
{\small 10} & {\small 9} & {\small 4.49$\,\cdot10^{-14}$} & {\small 0.01528} & {\small 1.67974} & {\small 1.86851} & {\small 1.50996} & {\small 8.17$\,\cdot10^{-13}$} & {\small 6.81$\,\cdot10^{-12}$} \\
\hline
{\small 20} & {\small 6} & {\small 4.95$\,\cdot10^{-14}$} & {\small 3.00$\,\cdot10^{-6}$} & {\small 1.61608} & {\small 2.02344} & {\small 1.75643} & {\small 6.50$\,\cdot10^{-12}$} & {\small 1.39$\,\cdot10^{-11}$} \\
\hline
{\small 100} & {\small 4} & {\small 2.52$\,\cdot10^{-14}$} & {\small 1.44$\,\cdot10^{-39}$} &
{\small 1.59316} & {\small 2.15456} & {\small 1.95689} & {\small 5.39$\,\cdot10^{-11}$} & {\small
  7.29$\,\cdot10^{-11}$} \\
\hline
{\highlight\small $\infty$} & {\highlight\small n.a.} & {\highlight\small n.a.} & {\highlight\small $0$} &
{\highlight\small 1.58927} & {\highlight\small 2.18738} & {\highlight\small 2.00705} &
{\highlight\small 1.26$\,\cdot10^{-10}$} & {\highlight\small 1.62$\,\cdot10^{-7}$}\\
\hline
\end{tabular}
\caption{Values of relevant parameters for $b\ge2$. For brevity we give only five decimal figures for
  the quantities of order 1, but we estimate our accuracy to be twice as much, {\highlight except perhaps for
  $b=\infty$}. $N_\mathrm{iter}$ is the number of iterations needed by the simple map
  \eqref{eq:simplemap} to satisfy the bound \eqref{eq:success}. After that, the minimization
  algorithm is run and $\cal R$ is the residual value when it stops. }
\label{table:table1}
\end{table}

\begin{figure}[tbp]
  \begin{center}
    \includegraphics[width=.75\textwidth]{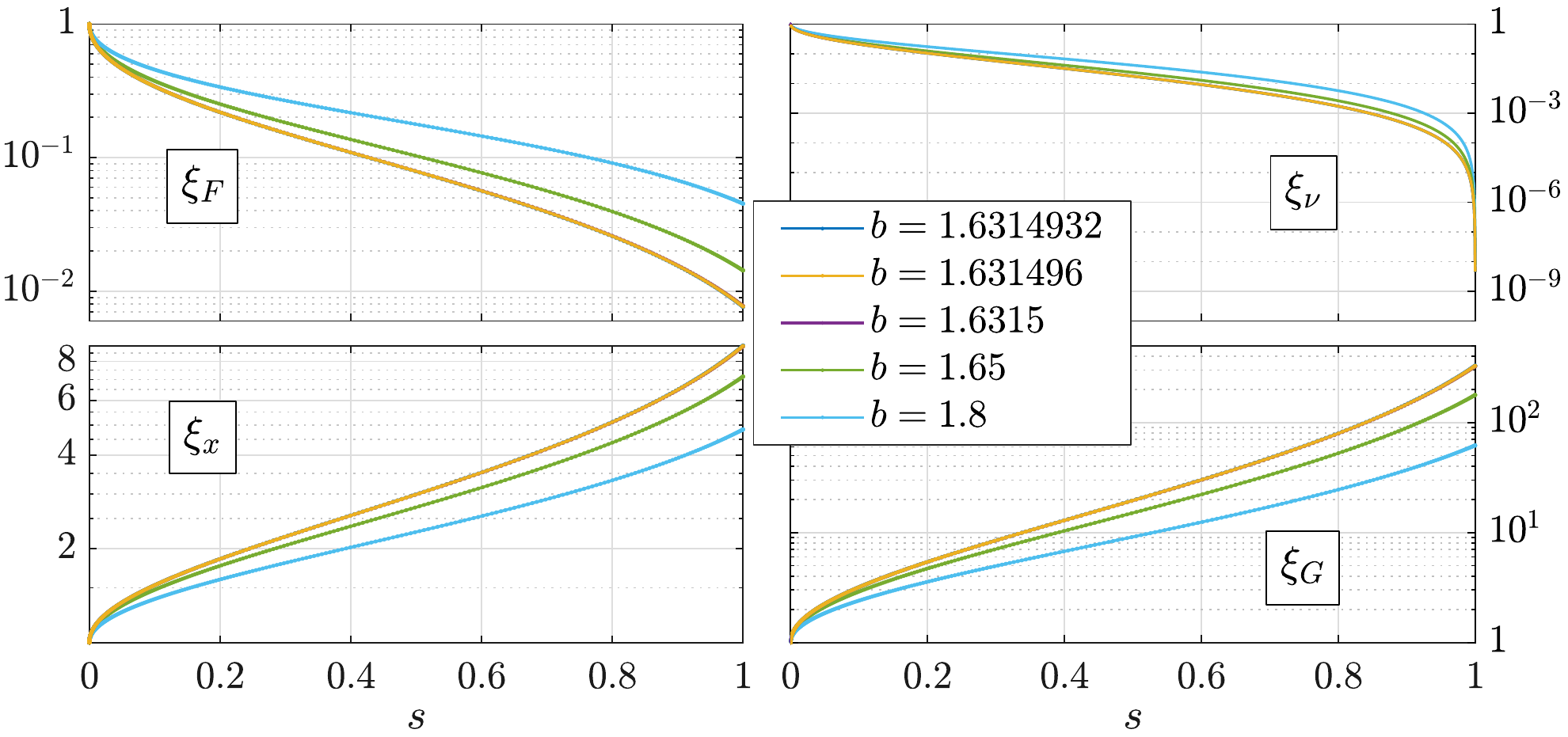}
    \caption{First kind $\xi_F$, $\xi_\nu$, $\xi_x$ and $\xi_G$ at selected values of $b<2$,
      with $h=0$ [see eqs.~\eqref{eq:F2xi}, \eqref{eq:G2xi} and \eqref{eq:nux2xi}. The profiles
      corresponding to the first three values of $b$ are indistinguishable on the scale of the plot.}
    \label{fig:profiles2}
  \end{center}
\end{figure}
\smallskip 

In the opposite direction, that is when $b$ is decreased, we find that $K$, $x(1)$, $m$ and $\eta$
monotonically increase. At the same time, the number of iterations of the map
eq.~\eqref{eq:simplemap} required to reach an accurate solution increases much more rapidly. A
significant time saving is then obtained by starting each new run from the final configuration
obtained in the preceding run, adapting the step down in $b$ to guarantee a reasonably fast
convergence. Some of the profiles are plotted in Fig \ref{fig:profiles2}, while the global data
shown in Table \ref{table:table1} for $b\ge2$ are now listed in Table \ref{table:table2} for few
selected values of $b$. One notices the steep rise of $K$, $x(1)$, $m$ and $\eta$ as $b$ approaches
the smallest value in the table, $b=1.6314932$. Repeating some of the runs with $N=1000$ Chebyshev
points just confirm the $N=400$ results.

\begin{table}
\centering
\begin{tabular}{|l|r|c|c|c|c|c|c|}
\hline
{\small $\quad\;\;b$} & {\small $N_\mathrm{iter}$} & {\small $\cal R$} & {\small $K$} & {\small $x(1)$} & {\small $m$} & {\small $\eta$} & {\small $\Delta m$} \\
\hline
{\small 1.6314932} & {\small 644496} & {\small 4.06$\,\cdot10^{-12}$} & {\small 5.45442} & {\small 12.71729} & {\small 2.82232} & {\small 0.95194} & {\small 1.14$\,\cdot10^{-11}$} \\
\hline
{\small 1.6314937} & {\small 180107} & {\small 5.70$\,\cdot10^{-12}$} & {\small 5.45239} & {\small 12.70294} & {\small 2.82127} & {\small 0.95190} & {\small 5.23$\,\cdot10^{-12}$} \\
\hline
{\small 1.631496} & {\small 84342} & {\small 7.08$\,\cdot10^{-12}$} & {\small 5.44927} & {\small 12.68105} & {\small 2.81966} & {\small 0.95184} & {\small 1.33$\,\cdot10^{-11}$} \\
\hline
{\small 1.63150} & {\small 21589} & {\small 2.09$\,\cdot10^{-12}$} & {\small 5.44625} & {\small 12.65990} & {\small 2.81810} & {\small 0.95178} & {\small 2.68$\,\cdot10^{-11}$} \\
\hline
{\small 1.63155} & {\small 6385} & {\small 6.83$\,\cdot10^{-12}$} & {\small 5.43031} & {\small 12.54914} & {\small 2.80991} & {\small 0.95147} & {\small 9.12$\,\cdot10^{-12}$} \\
\hline
{\small 1.65} & {\small 456} & {\small 5.30$\,\cdot10^{-13}$} & {\small 5.01071} & {\small 10.11928} & {\small 2.61357} & {\small 0.94383} & {\small 1.25$\,\cdot10^{-11}$} \\
\hline
{\small 1.8} & {\small 150} & {\small 5.49$\,\cdot10^{-13}$} & {\small 4.10702} & {\small 6.84120} & {\small 2.28767} & {\small 0.93489} & {\small 9.53$\,\cdot10^{-12}$} \\
\hline
\end{tabular}
\caption{Values of relevant parameters for $b<2$. $N_\mathrm{iter}$ is the cumulative number of
  iterations needed to satisfy the bound \eqref{eq:success}. For lack of space we do not list the values of
  $\Delta\eta$, which anyway are of the same order of  $\Delta m$.}
\label{table:table2}
\end{table}

Clearly the solutions just listed for $b\ge2$ and for $1.6314932\le b<2$ belong to a continuous
family, which we call of the first kind, with the prominent feature that
\begin{equation}\label{eq:down}
  \xi_F(s,b) < \xi_F(s,b') \;,\quad \forall s\in[0,1]\;,\quad\mathrm{if}\; b < b'\;.
\end{equation}
As a consequence, $\xi_\nu$ also increases with $b$ while $\xi_x$ and $\xi_G$ uniformly decrease
[this is due to the positivity property highlighted below eq.~\eqref{eq:intpois}]. Thus 
$K$ and $x(1)$ are both monotonically decreasing with $b$. The behavior of the $m$ and $\eta$ in
the interval $1.6314932\le b\le 10$ is not monotonic, but we suspect $\eta$ to be monotonically increasing
for all $b>2$.

\smallskip 
At each solution found, it is of interest to compute through finite differences a good
approximation of the Jacobian matrix $J(\alpha)$, eq.~\eqref{eq:jac}, associated to the iterated map
\eqref{eq:simplemap}.  We find that all eigenvalues of $J(1)$ have magnitude smaller than 1, as
expected since the iterated map always converges. However the largest eigenvalue, which is real,
approaches unity as $b\to 1.6314932$, causing the observed slow--down.

\begin{figure}[tbp]
  \begin{center}
    \includegraphics[width=.75\textwidth]{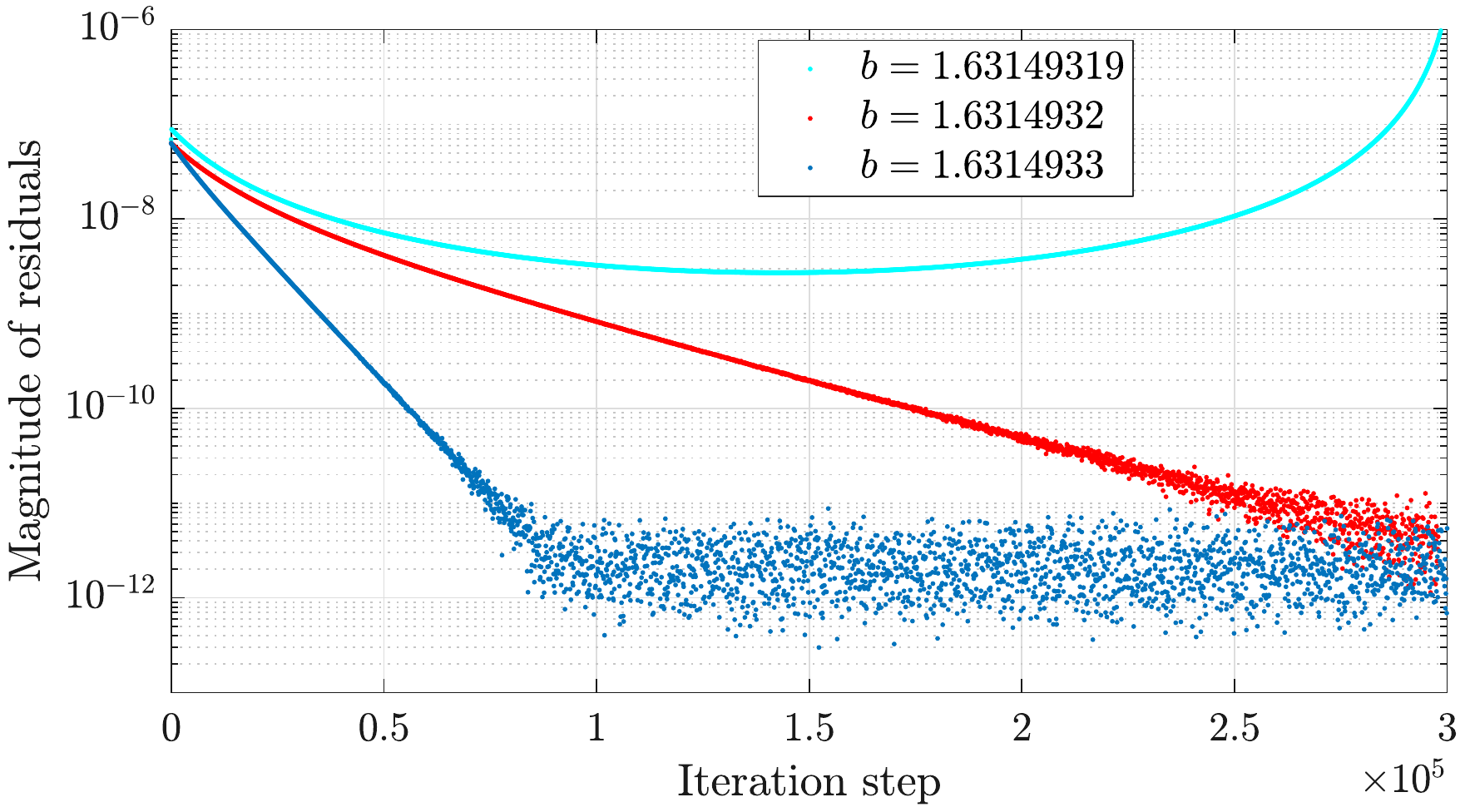}
    \caption{The evolution of $\cal R$, the largest residual magnitude, along the iterated map
      \eqref{eq:simplemap} with comparable initial conditions, for three very close values of $b$.}
    \label{fig:simplemap3}
  \end{center}
\end{figure}
\smallskip 

%\medskip 
We find it impossible to go below $b=1.6314932$ by any amounts of order $10^{-8}$ or more, while
preserving the default smallness of the residuals, that is ${\cal R}<10^{11}$. In
Fig.~\ref{fig:simplemap3} we plot the residual $\cal R$ during three runs of the iterated map
\eqref{eq:simplemap}, when $b$ takes the values $b=1.6314933$, $b=1.6314932$ and $b=1.63149319$. The
first run starts at the $\xi_F$ that satisfies eq.~\eqref{eq:xixi}, in the sense of criterion
\eqref{eq:success}, when $b=1.6314934$; the second and the third start both at the solution
corresponding to $b=1.6314933$, when we still obtain a successful run, as evident from
Fig.~\ref{fig:simplemap3}. Notice that, by construction, the initial $\cal R$ is slightly less than
$10^{-7}$ in all three cases. While the iteration with $b=1.6314932$ still manages to converge,
albeit very slowly, the iteration with $b=1.63149319$, after a decrease to values of $\cal R$ of
order $10^{-9}$, eventually begins to diverge. Since we expect the solutions to depend continuously on
$b$, we infer that at a certain value $b=b_1$ between $1.63149319$ and $b=1.6314932$ two
solutions of eq.~\eqref{eq:xixi} that existed for $b>b_1$ have coalesced and there are no
solutions at all for $b\lesssim b_1$

Up to now, however, we have only found one solution for each value of $b>b_1$, either starting
from $\xi_F^{(0)}(s)=1$ or relying on continuity in $b$ and using as starting point of a new run the
final profile of a successful preceding run. This means that, when descending from $b=2$, the final
profile of $\xi_F$ is uniformly smaller than the initial one [see eq.~\eqref{eq:down}]. In this way
we might have missed the possibility of even smaller $\xi_F$ solutions. Indeed, in the next
subsection we describe such other solution family for $b>b_1$.

\subsection{Solutions of the second kind}\label{sec:secondkind}

\begin{figure}[tbp]
  \begin{center}
    \includegraphics[width=.8\textwidth]{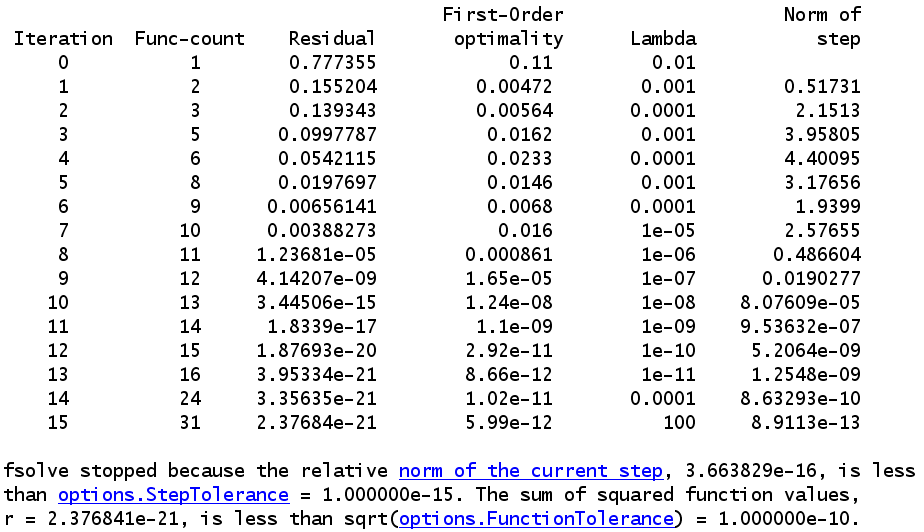}
    \caption{Output of MATLAB {\tt fsolve} running the Levenberg--Marquardt algorithm when $b=1.8$ and
      $N=400$, starting from the $\xi_F$ profile obtained for $b=1.6314932$.}
    \label{fig:iter_val2}
  \end{center}
\end{figure}

\begin{figure}[tbp]
  \begin{center}
    \includegraphics[width=.68\textwidth]{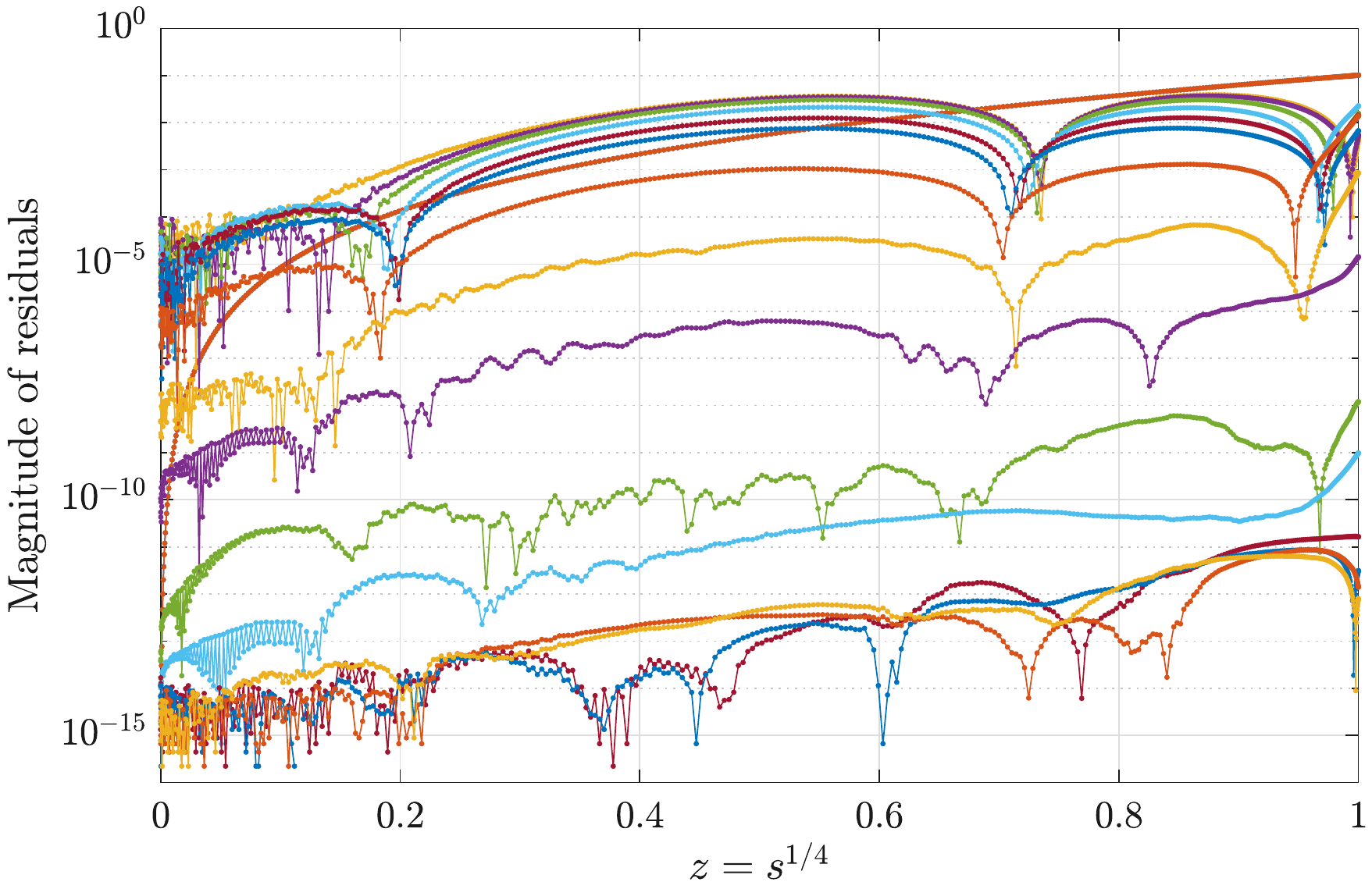}
    \caption{The absolute values of the residuals $R_k$ at each
      iteration of {\tt fsolve} running as in Fig~\ref{fig:iter_val2}.}
    \label{fig:iter_resid2}
  \end{center}
\end{figure}

Suppose we use the final $\xi_F$ found for $b=1.6314932$, the value closest to the boundary value
$b_1$, when we still obtain an accurate solution, as starting point for a new run with a
\emph{larger} value of $b$, say the next value in Table \ref{table:table2}, $b=1.6314937$. This
$\xi_F$ is uniformly smaller than the one we have already found as accurate solution to
eq.~\eqref{eq:xixi} when $b=1.6314937$ and, correspondingly, $K(1.6314932)>K(1.6314937)$. We find
that the iterated map \eqref{eq:simplemap} just reproduces that solution, while the minimization
algorithm fails to reach the desired accuracy, even playing around with its many tunable parameters,
in any reasonable amount of time. Within our hypothesis of the coalescence of two roots, this
probably means that $b=1.6314937$ is still too close to coalescence value $b_1$, for the
minimization algorithm to disentangle the new solution we are seeking for from the nearby solution
we have already found.

However, since at first we only need to find the new solution for just one value of $b$, let us
consider a larger value of $b$, far enough from the coalescence region. Let us pick $b=1.8$, the
largest value in table \ref{table:table1}, as starting value for the sought second branch of
solutions. We discover that both methods to solve eq.~\eqref{eq:xixi} are successful, but yield
different results. While in 152 iterations the iterated map \eqref{eq:simplemap} with $\alpha=1$
falls back to the solution with $K=4.107017\ldots$ and $x(1)=6.841203\ldots$, already found and
listed in Table \ref{table:table2}, the minimization algorithm, just in its simplest implementation,
converges [see Figs.~\ref{fig:iter_val2} and~\ref{fig:iter_resid2}] to a new profile, uniformly
smaller than $\xi_F(s,b_1)$ and therefore with the larger normalization factor $K=6.692666\ldots$
and a larger size $x(1)=35.119263\ldots$.

\begin{table}
\centering
\begin{tabular}{|l|c|c|c|c|c|c|c|}
\hline
{\small $\quad\;\; b$} & {\small $\cal R$} & {\small $K$} & {\small $x(1)$} & {\small $m$} & {\small $\eta$} & {\small $\Delta m$} & {\small $\Delta\eta$} \\
\hline
{\small 1.6314937} & {\small 9.23$\,\cdot10^{-12}$} & {\small 5.45699} & {\small 12.73540} & {\small 2.82365} & {\small 0.95199} & {\small 2.96$\,\cdot10^{-12}$} & {\small 6.25$\,\cdot10^{-13}$} \\
\hline
{\small 1.631496} & {\small 1.47$\,\cdot10^{-12}$} & {\small 5.46010} & {\small 12.75743} & {\small 2.82526} & {\small 0.95205} & {\small 2.62$\,\cdot10^{-11}$} & {\small 4.67$\,\cdot10^{-12}$} \\
\hline
{\small 1.63150} & {\small 8.16$\,\cdot10^{-12}$} & {\small 5.46312} & {\small 12.77884} & {\small 2.82683} & {\small 0.95211} & {\small 2.40$\,\cdot10^{-11}$} & {\small 8.22$\,\cdot10^{-12}$} \\
\hline
{\small 1.63155} & {\small 7.97$\,\cdot10^{-13}$} & {\small 5.47903} & {\small 12.89281} & {\small 2.83512} & {\small 0.95242} & {\small 2.51$\,\cdot10^{-11}$} & {\small 6.31$\,\cdot10^{-12}$} \\
\hline
{\small 1.65} & {\small 6.39$\,\cdot10^{-13}$} & {\small 5.88779} & {\small 16.55156} & {\small 3.07105} & {\small 0.96013} & {\small 3.39$\,\cdot10^{-11}$} & {\small 7.05$\,\cdot10^{-12}$} \\
\hline
{\small 1.8} & {\small 6.39$\,\cdot10^{-12}$} & {\small 6.69267} & {\small 35.11926} & {\small 3.72791} & {\small 0.95621} & {\small 7.29$\,\cdot10^{-11}$} & {\small 6.63$\,\cdot10^{-12}$} \\
\hline
\end{tabular}
\caption{Values of relevant parameters for $b\le1.8$ in the family of solutions of the second kind.}
\label{table:table3}
\end{table}

Having obtained a new, or second--kind, solution for $b=1.8$, we can now proceed and progressively
reduce $b$ toward $b_1$, using as starting point of a new run the final profile of the preceding
successful run. This should imply a uniform increase in $\xi_F(s)$, with a corresponding decrease in
$K$, but always remaining above $K(b_1)$, the value of $K$ that separates the two
branches. Albeit slow, this procedure indeed works as expected, with the results listed in Table
\ref{table:table3} (it is actually too slow to reach the value $b=1.6314932$, probably because the
other first--kind solution is too close). Notice that only the minimization algorithm works
successfully, since the iterated map always falls back to a solution of the first kind with
$K<K(b_1)$. In other words the second--kind solution at a given $b$ represents an unstable fixed
point of the iterated map \eqref{eq:simplemap}. We can verify this by looking at the eigenvalues of
the Jacobian matrix $J(1)$ in eq.~\eqref{eq:jac}. We find that the largest eigenvalue now exceeds
one, while there are eigenvalues with negative real part. This means that no choice of $\alpha$
would make the iterated map \eqref{eq:simplemap} convergent.

\begin{table}
\centering
\begin{tabular}{|l|c|c|c|c|c|c|c|}
\hline
{\small $\;\;b$} & {\small $\cal R$} & {\small $K$} & {\small $x(1)$} & {\small $m$} & {\small $\eta$} & {\small $\Delta m$} & {\small $\Delta\eta$} \\
\hline
{\small 1.9} & {\small 3.77$\,\cdot10^{-12}$} & {\small 6.96731} & {\small 59.91614} & {\small
  4.05790} & {\small 0.92821} & {\small 2.38$\,\cdot10^{-10}$} & {\small 9.47$\,\cdot10^{-12}$} \\
\hline
{\small 1.96} & {\small 8.68$\,\cdot10^{-12}$} & {\small 7.09661} & {\small 96.16559} & {\small 4.24662} & {\small 0.89918} & {\small 3.98$\,\cdot10^{-10}$} & {\small 1.13$\,\cdot10^{-11}$} \\
\hline 
{\small 2} & {\small 7.01$\,\cdot10^{-12}$} & {\small 7.17359} & {\small 159.27710} & {\small
  4.37142} & {\small 0.87314} & {\small 9.85$\,\cdot10^{-10}$} & {\small 3.57$\,\cdot10^{-11}$} \\
\hline\hline
{\small 2.01} & {\small 8.46$\,\cdot10^{-12}$} & {\small 7.19204} & {\small 190.69520} & {\small 4.40270} & {\small 0.86565} & {\small 8.81$\,\cdot10^{-11}$} & {\small 1.12$\,\cdot10^{-10}$} \\
\hline
{\small 2.02} & {\small 9.30$\,\cdot10^{-12}$} & {\small 7.21025} & {\small 237.78943} & {\small 4.43405} & {\small 0.85773} & {\small 6.09$\,\cdot10^{-10}$} & {\small 2.09$\,\cdot10^{-11}$} \\
\hline
{\small 2.03} & {\small 3.76$\,\cdot10^{-11}$} & {\small 7.22827} & {\small 316.28491} & {\small 4.46550} & {\small 0.84937} & {\small 1.27$\,\cdot10^{-09}$} & {\small 5.64$\,\cdot10^{-10}$} \\
\hline
{\small 2.04} & {\small 8.83$\,\cdot10^{-11}$} & {\small 7.24614} & {\small 473.44293} & {\small 4.49709} & {\small 0.84056} & {\small 5.24$\,\cdot10^{-09}$} & {\small 7.52$\,\cdot10^{-10}$} \\
\hline
{\small 2.05} & {\small 8.71$\,\cdot10^{-10}$} & {\small 7.26391} & {\small 946.52061} & {\small 4.52884} & {\small 0.83129} & {\small 5.46$\,\cdot10^{-08}$} & {\small 5.07$\,\cdot10^{-09}$} \\
\hline
{\small 2.055} & {\small 8.08$\,\cdot10^{-09}$} & {\small 7.27277} & {\small 1899.810} & {\small 4.54478} & {\small 0.82648} & {\small 8.54$\,\cdot10^{-07}$} & {\small 3.93$\,\cdot10^{-08}$} \\
\hline
\end{tabular}
\caption{Values of relevant parameters for $b>1.8$ in second--kind solutions. When $b\le2(b>2)$ the
  calculations were performed using $400(10^3)$ Chebyshev points. }
\label{table:table4}
\end{table}

\begin{figure}[tbp]
  \begin{center}
    \includegraphics[width=.75\textwidth]{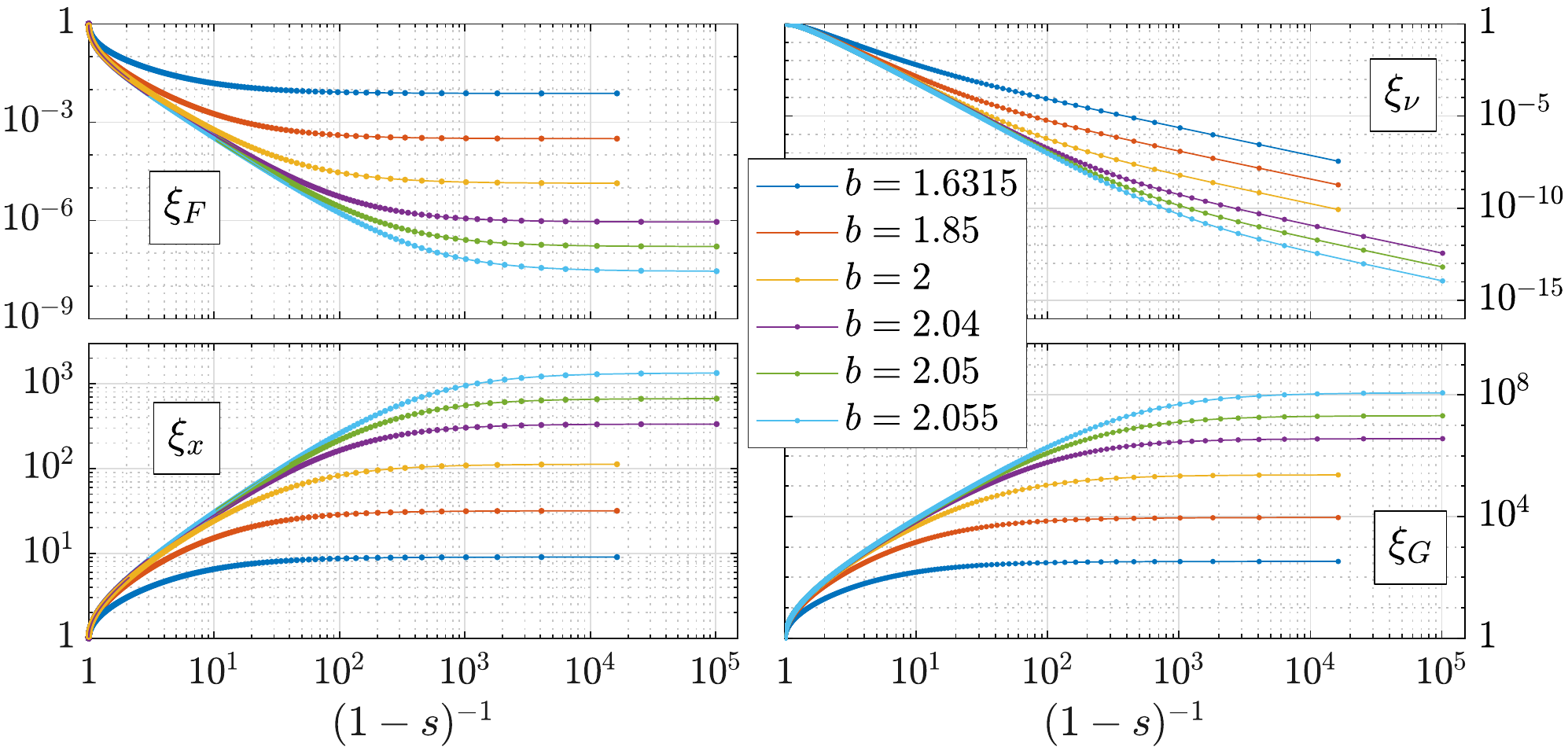}
    \caption{Second--kind $\xi_F$, $\xi_\nu$, $\xi_x$ and $\xi_G$ at selected values of $b$. These
      are profiles with $h=0$ in eqs.~\eqref{eq:F2xi}, \eqref{eq:G2xi} and \eqref{eq:nux2xi},
      although in the actual calculations $h$ was set to values growing with $b$ towards $1$, to
      absorb as much as possible the large growth of $\xi_x$ and $\xi_G$. The profiles corresponding
      to the three smaller values of $b$ are defined over the grid with $N=400$ points, while the other
      three are defined over the grid with $N=1000$. }
    \label{fig:profiles3}
  \end{center}
\end{figure}

\smallskip To try and complete the branch of solutions of the second kind, we next progressively
increase $b$ from the value $1.8$. Now $K$ and $x(1)$ increase with $b$, with the size $x(1)$
increasing very fast for $b\gtrsim2$. The rise of $x(s)$ as $s\to1$ becomes so steep that it can
compromise the accuracy of the calculation of the integral transform \eqref{eq:xix2xiG} when using
$N=400$ Chebyshev points, even if the free parameter $h$ is tuned to absorb as much as possible the
growth in the prefactor $(1-h s)^{-1}$, according to eqs.~\eqref{eq:nux2xi}. Notice that one cannot
move $h$ too close to unity, or $\xi_F(s)$ would develop a steep rise as $s\to1$, which would in
turn lower the quality of the Chebyshev approximation of the other integral
transform~\eqref{eq:xiF2xinu}. For this reason we use also $N=1000$ Chebyshev points for all values
$b>2$ where we search for solutions.
 
The results for some specific values of $b$ are listed in Table \ref{table:table4}, while some of
the profiles are plotted in Fig.~\ref{fig:profiles3}. Evidently, for $b\gtrsim2$ the accuracy
progressively worsen. As discussed in the Appendix, this is due to the second step in the chain
\eqref{eq:chain}, that is Poisson's equation \eqref{eq:pois2}, because of the increasing span of its
solution $x(u)$. Although the bound \eqref{eq:success} on the residuals is not satisfied for the
last three entries of the table, relying on the continuity in $b$ we can still regard them as
adequate, if not fully accurate, numerical solutions.

The solutions of the second kind belong to a family continuous in $b$, with the prominent feature that
\begin{equation}\label{eq:up}
  \xi_F(s,b) > \xi_F(s,b') \;,\quad \forall s\in[0,1]\;,\quad\mathrm{if}\; b < b'\;,
\end{equation}
which is a behavior opposite w.r.t. the first kind [see eq.~\eqref{eq:down}]. Another, even more
prominent feature of this second--kind family is the existence of a value $b=b_2$ where the
dimensionless size $x(1)$ diverges. We can find an approximate value of $b_2$ by fitting $1/x(1)$
vs. $b$. Using the six values of $x(1)$ listed in Table \ref{table:table4} for $b>2$ (obtained with
the more accurate setup based on $N=1000$ Chebyshev points) we consider the polynomial
interpolations of $1/x(1)$ vs. $b$ from degree one to five (with centering and scaling for the
higher degrees). The root of each polynomial which is closest to $b=2.055$ provides an approximate
value of $b_2$. We find
\begin{equation*}
\centering
\begin{tabular}{|c|c|c|c|c|c|}
\hline
{degree of polynomial} & {\small $b_2$} & {max residual}  \\
\hline
{\small 1} & {\small 2.0600857} & {\small 1.418$\,\cdot10^{-5}$} \\
\hline
{\small 2} & {\small 2.0599539} & {\small 3.975$\,\cdot10^{-7}$} \\
\hline
{\small 3} & {\small 2.0599476} & {\small 9.948$\,\cdot10^{-8}$} \\
\hline
{\small 4} & {\small 2.0599507} & {\small 6.046$\,\cdot10^{-9}$} \\
\hline
{\small 5} & {\small 2.0599511} & {\small 1.697$\,\cdot10^{-18}$} \\
\hline
\end{tabular}
\end{equation*}
This calculation is stable and shows that $x(1)$ diverges as $(b_2-b)^{-1}$ as $b\to b_2$. Moreover,
the profiles in Fig.~\ref{fig:profiles3} show the emergence, as $b\to b_2$, of scaling windows where 
$\xi_F$, $\xi_\nu$, $\xi_x$ and $\xi_G$ have  near $s=1$ the power behavior characteristic of infinite--size
systems, as discussed in subsection \ref{sec:infinite}. 

We do not know whether the second--kind family continues beyond $b_2$ with $x(1)$ infinite but
$b-$dependent mass and temperature. Unfortunately, we cannot solve Poisson's equation
\eqref{eq:pois2} with the necessary accuracy when $\lim_{u\to1}x(u)=\infty$.

\subsection{A unifying view of the two kinds of solutions}

\begin{figure}[tbp]
  \begin{center}
    \includegraphics[width=.75\textwidth]{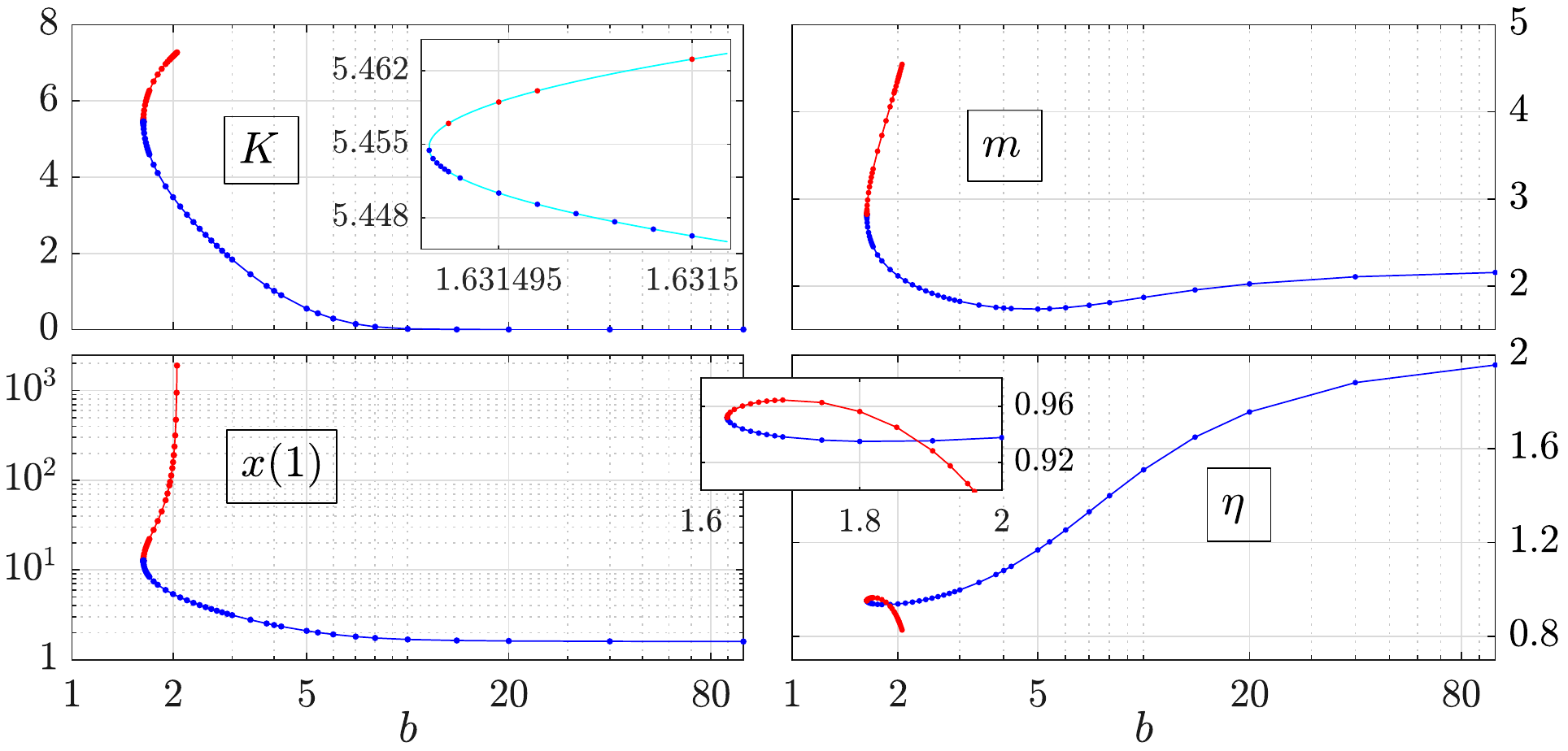}
    \caption{Plot of $K$, $x(1)$, $m$ and $\eta$ vs. $b$. The dots (blue for first--kind solutions
      and red for second--kind solutions) mark where the calculation was performed. The continuous
      cyan curve in the inset of the upper right panel is the parabolic fit to the actual data
      marked by the dots. }
    \label{fig:Kmxeta_vsb}
  \end{center}
\end{figure}

In Fig.~\ref{fig:Kmxeta_vsb} we plot $K$, $x(1)$, $m$ and $\eta$ vs. $b$, considering both families
of solutions. Obviously, for $b_1<b<b_2$ these quantities are all double--valued as functions of
$b$. Viceversa, $b$ is a single--valued function of $K$ and of $x(1)$, while it is double--valued as
function of $m$ or of $\eta$. Near the lowest value $b_1$, say for $b\le 1.6315$, $b$ itself is
very well approximated, as a function of $K$, $x(1)$, $m$ or $\eta$, by a parabola. In the inset in
the upper left panel of Fig.~\ref{fig:Kmxeta_vsb} we show this fit in the case of $b$ vs. $K$. These
fits allow accurate approximations for $b_1$ as the minimum of each parabola. Actually, better
accuracy is obtained by using third order polynomials rather than parabolas (clearly $b$ is not even
as function of $K$, $x(1)$, $m$ or $\eta$). We find
\begin{equation*}
\centering
\begin{tabular}{|c|c|c|c|c|}
\hline
{\small variable} & {\small $b_1$} & {\small value at $b_1$} &{\small max residual}  \\
\hline
{\small $K$} & {\small 1.63149319330} & {\small 5.45468873434} & {\small 4.14$\,\cdot10^{-11}$} \\
\hline
{\small $x(1)$} & {\small 1.63149319327} & {\small 12.7191519275} & {\small 1.176$\,\cdot10^{-10}$} \\
\hline
{\small $m$} & {\small 1.63149319329} & {\small 2.82245779594}& {\small 5.14$\,\cdot10^{-11}$} \\
\hline
{\small $\eta$} & {\small 1.63149319328} & {\small 0.95194462704} & {\small 1.093$\,\cdot10^{-10}$} \\
\hline
\end{tabular}
\end{equation*}
Taking into account that $10^{-11}$ is the accuracy of both solution families, we conclude that 
\begin{equation}
  b_1 = 1.631493193(3) \;.
\end{equation}

\begin{figure}[tbp]
  \begin{center}
    \includegraphics[width=.75\textwidth]{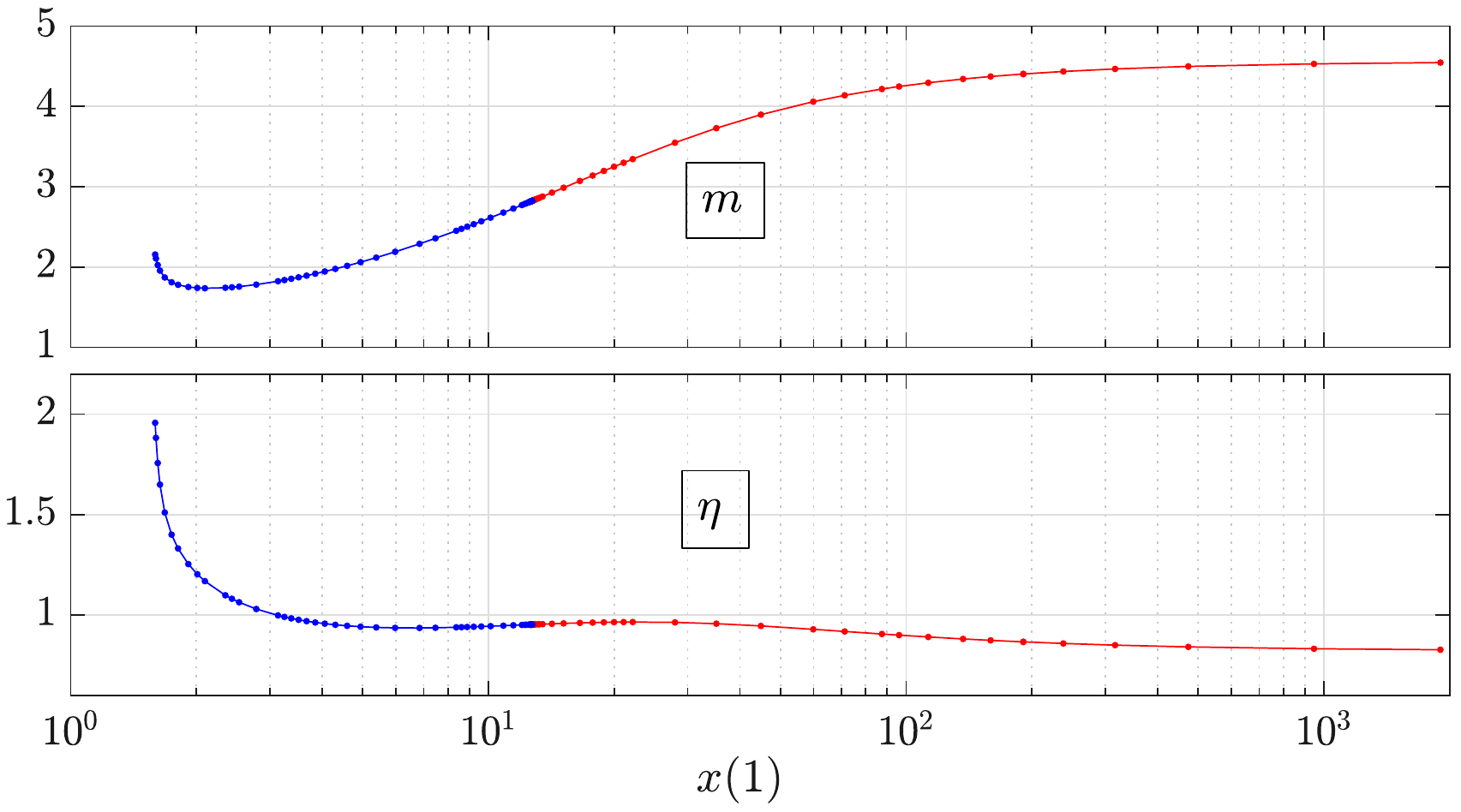}
    \caption{Plot of $m$ and $\eta$ vs. $x(1)$. Again, blue(red) dots mark actual calculations for
      solutions of the first (second) kind.}
    \label{fig:meta_vs_xe}
  \end{center}
\end{figure}

The parameter $b$ is just a Lagrange multiplier and, after computing the physical observables, it
can be eliminated in favor of one such observable; the dimensionless size $x(1)$ is clearly the best
choice. In Fig.~\ref{fig:meta_vs_xe} we plot the dimensionless mass $m$ and temperature $\eta$
vs. $x(1)$. No particular feature highlights the transition from the family of the first kind to the
second--kind family. Such transition is evident only because of the different colors used in the plots.
Similarly, the four profiles $\xi_F$, $\xi_\nu$, $\xi_x$ and $\xi_G$ have a smooth and regular
dependence on $x(1)$: $\xi_F$ and $\xi_\nu$ monotonically and uniformly decrease as $x(1)$
grows, while $\xi_x$ and $\xi_G$ monotonically and uniformly increase.

\begin{figure}[tbp]
  \begin{center}
    \includegraphics[width=.75\textwidth]{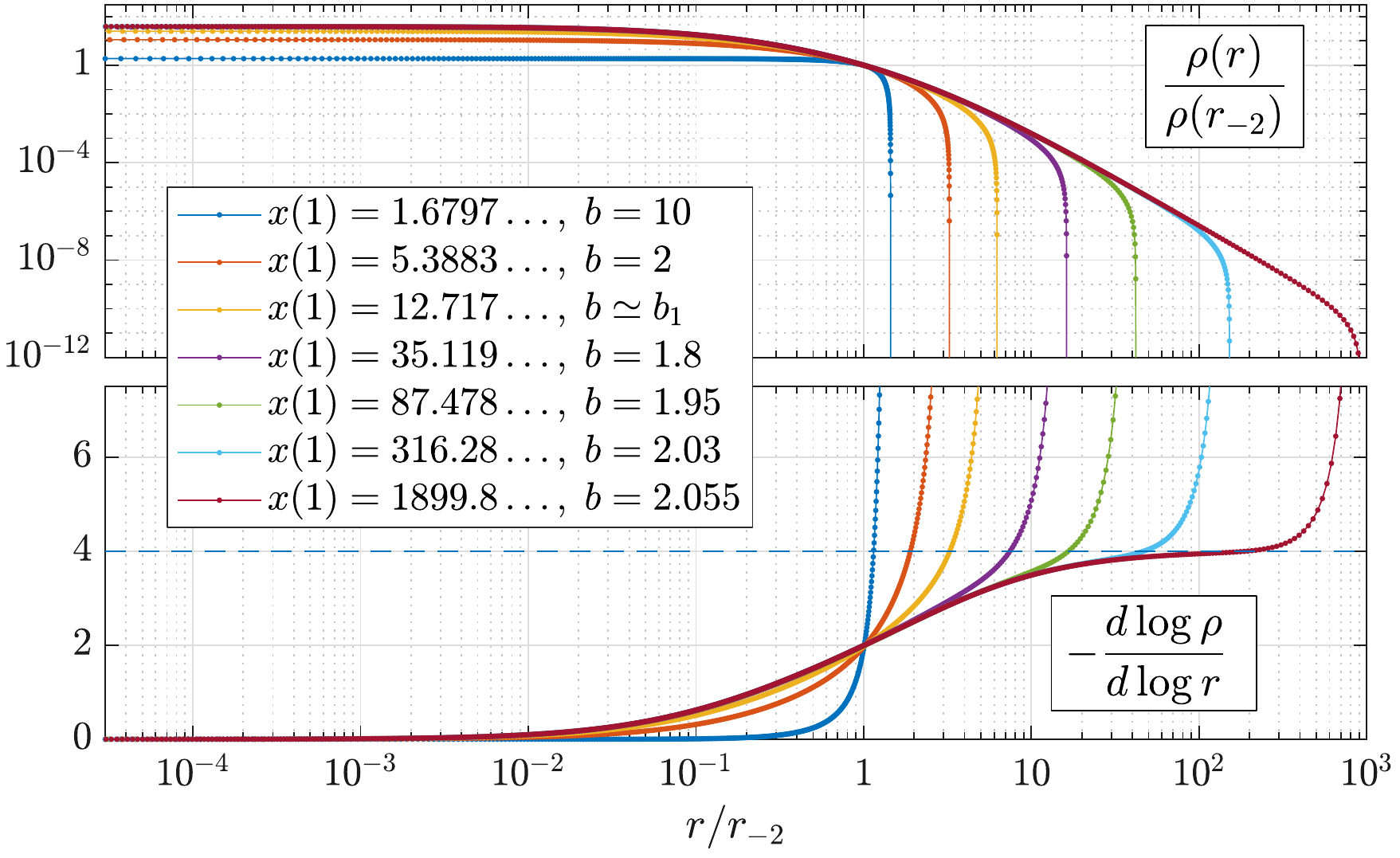}
    \caption{The density profiles (upper panel) and their logarithmic slopes (lower panel) for few
      selected solutions; the first three are from the family of the first kind and the last four
      from that of the second kind.}
    \label{fig:rho_slopes}
  \end{center}
\end{figure}

A very important profile is that of $\rho(r)$, the mass density vs. the radius. In Section
\ref{sec:dless}, to fix the distance scale we used $r_\ast$, which relates the central density
$\rho_0$ (a local quantity) to the total mass $M$ (a global quantity) as in
eq.~\eqref{eq:M2m}. Another, fully local scale, denoted $r_{-2}$, is often used. It is defined as
the largest radius at which the logarithmic slope $r\rho'(r)/\rho(r)$ takes the value $-2$. In the
upper panel of Fig.~\ref{fig:rho_slopes} we plot several profiles of $\rho(r)/\rho(r_{-2})$
vs. $r/r_{-2}$. Again, nothing special happens at the transition from the first-- to the
second--kind family. Notice also the emergence of the $r^{-4}$ law at large distances when $b$
approaches $b_2$, as evident also from the logarithmic slopes plotted in the lower panel.

\begin{figure}[tbp]
  \begin{center}
    \includegraphics[width=.75\textwidth]{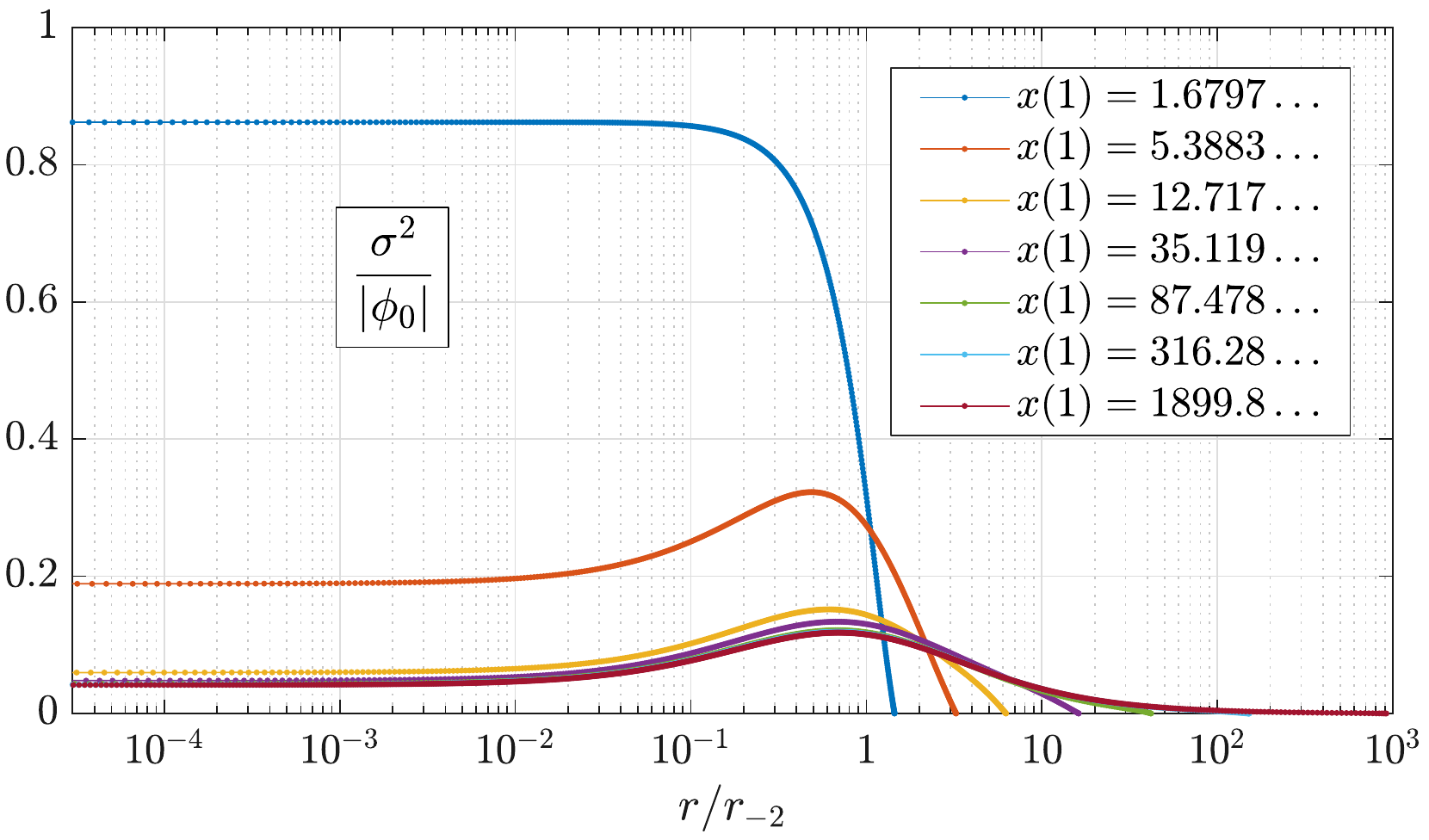}
    \caption{The squared velocity dispersion for the same selected solutions of
      Fig.~\ref{fig:rho_slopes}.}
    \label{fig:sigmassq}
  \end{center}
\end{figure}

Finally, in Fig.~\ref{fig:sigmassq} we plot the squared velocity dispersion [see
eqs.~\eqref{eq:sigma} and ~\eqref{eq:sigma1}] for the same cases of Fig.~\ref{fig:rho_slopes}. It is
evident as these cored DARKexp systems get cooler as their size grows.

\subsection{Asymptotic fit}\label{sec:afit}

In the Appendix we present an apriori analysis on the accuracy of our numerical results, with a
typical estimate of order $10^{-11}$. In this section we check the numerical accuracy against the
only exact analytical result that we know, that is the next--to--leading asymptotics
\eqref{eq:nnlasy} derived in \cite{destri2018}.

In principle, having computed the Chebyshev approximation \eqref{eq:chebexp} for the four functions
$\xi_F$, $\xi_\nu$, $\xi_x$ and $\xi_G$, to compute the first correction of order $u^{-1/2}$ in
eq.~\eqref{eq:nnlasy} we could just evaluate the second derivative of $\xi(z^4)$ w.r.t. $z$ at
$z=0$. This procedure is however quite inaccurate, due to the intrinsic nature of polynomial
interpolations, which in our case in aggravated by the expected presence of a singular logarithmic
correction. A better procedure, which allows to estimate also the coefficient of logarithmic
correction, is to use a finite version of the expansion \eqref{eq:asympt} to fit our numerical
$\xi(s)$ over a small subinterval $[0,d]$, with $d\ll1$ suitably chosen to maximize the reliability
of the fit. For small enough $d$ only relatively few coefficients in the expansion are numerically
relevant and any non--linear least--squares method [for instance the Levenberg--Marquardt algorithm
already at work with eq.~\eqref{eq:xixi}] converges very quickly. Initially we can pick $d$ so small
that $N_d$, the number of grid points in $[0,d]$ (that is the number of values to be fitted), is not
much larger than $M_d$, the number of relevant coefficients. Then $d$ could be increased as long as
the fit improves, reducing the least--square residuals multiplied by the ratio $M_d/N_d$.

An obvious limitation applies to the very compact solutions where $K\ll1$, since the numbers we
want to check are, for instance in the case of $\xi=\xi_F$,
\begin{equation}\label{eq:bFaF}
  b_F = -\frac{45b_{F,00}}{32K} \;,\quad a_F=-\frac{45a_{F,10}}{2K^2}\;.
\end{equation}
According to eq.~\eqref{eq:nnlasy} these should both take the value $1$ in any solution,
irrespective of $b$ and of the family kind. Clearly accuracy is lost when $K\ll1$ and $b_F$, $a_F$
are ratios of very small numbers.

\begin{table}
\centering
\begin{tabular}{|c|r|c|c|c|c|l|c|}
\hline
{\small $b$} & {\small $x(1)$} & {\small $d$} & {\small $M_d$} & {\small $N_d$} & {\small $R M_d/N_d$} & {\small $\;\;|1-b_F|$} & {\small $|1-a_F|$} \\
\hline
{\small 5} & {\small 2.096} & {\small 3.2$\,\cdot10^{-7}$} & {\small 8} & {\small 40} & {\small 3.69$\,\cdot10^{-13}$} & {\small 8.87$\,\cdot10^{-9}$} & {\small 2.77$\,\cdot10^{-3}$} \\
\hline
{\small 2} & {\small 5.388} & {\small 3.2$\,\cdot10^{-7}$} & {\small 8} & {\small 40} & {\small 5.25$\,\cdot10^{-14}$} & {\small 3.82$\,\cdot10^{-9}$} & {\small 1.93$\,\cdot10^{-4}$} \\
\hline
{\small 1.6315} & {\small 12.660} & {\small 8.0$\,\cdot10^{-8}$} & {\small 8} & {\small 34} & {\small 1.16$\,\cdot10^{-14}$} & {\small 1.13$\,\cdot10^{-9}$} & {\small 7.06$\,\cdot10^{-5}$} \\
\hline
{\small $b_1$} & {\small 12.717} & {\small 8.0$\,\cdot10^{-8}$} & {\small 8} & {\small 34} & {\small 1.14$\,\cdot10^{-14}$} & {\small 9.39$\,\cdot10^{-10}$} & {\small 6.32$\,\cdot10^{-5}$} \\
\hline
{\small 1.6315} & {\small 12.779} & {\small 1.6$\,\cdot10^{-7}$} & {\small 8} & {\small 37} & {\small 4.98$\,\cdot10^{-14}$} & {\small 3.05$\,\cdot10^{-9}$} & {\small 9.58$\,\cdot10^{-5}$} \\
\hline
{\small 2} & {\small 159.277} & {\small 8.0$\,\cdot10^{-8}$} & {\small 8} & {\small 34} & {\small 4.85$\,\cdot10^{-14}$} & {\small 1.14$\,\cdot10^{-9}$} & {\small 6.71$\,\cdot10^{-5}$} \\
\hline
{\small 2.03} & {\small 316.285} & {\small 3.2$\,\cdot10^{-7}$} & {\small 8} & {\small 99} & {\small 1.25$\,\cdot10^{-13}$} & {\small 6.92$\,\cdot10^{-9}$} & {\small 1.68$\,\cdot10^{-4}$} \\
\hline
{\small 2.05} & {\small 946.521} & {\small 3.2$\,\cdot10^{-7}$} & {\small 8} & {\small 99} & {\small 1.29$\,\cdot10^{-13}$} & {\small 7.08$\,\cdot10^{-9}$} & {\small 1.69$\,\cdot10^{-4}$} \\
\hline
{\small 2.055} & {\small 1899.811} & {\small 5.1$\,\cdot10^{-6}$} & {\small 8} & {\small 141} & {\small 1.88$\,\cdot10^{-10}$} & {\small 1.87$\,\cdot10^{-6}$} & {\small 4.14$\,\cdot10^{-3}$} \\
\hline
\end{tabular}
\caption{Asymtotic fit of $\xi_F(s)$ to eq.~\eqref{eq:asympt}, with numerical check of the exact
  relations in eq.~\eqref{eq:bFaF}. }
\label{table:table5}
\end{table}

The results of our asymptotic fit are listed in Table \ref{table:table5} for a selected set of
numerical solutions. We only consider $a_F$ and $b_F$, since the other three cases, when $F$ is
replaced by $\nu$, $x$ or $G$, yield very similar results. Keeping into account the intrinsic
weakness of this kind of fit, the agreement of numbers vs. analytics is extremely good for
$b_F$. The lower level of agreement for $a_F$ is to be expected, due to the singular nature
of the log. 

\section{Discussion and outlook}

In this work we have numerically computed very accurate solutions of the DARKexp problem, namely, we
calculated ergodic phase--space distribution functions $f(E)$ that determine the DARKexp
differential energy distribution $n(E)$ given by eq.~\eqref{eq:nofe}. We have assumed cored density
profiles, by imposing the asymptotic condition $f(E)\sim (E-\phi_0)^{-1}$ near the bottom of the
potential well. As a consequence, beside $\phi_0$, we have the density $\rho_0$ in the origin as
second free parameter fixing the two scales of the problem. Together with the choice $4\pi G=1$,
we thus obtain a fully dimensionless formulation, described in section \ref{sec:dless}, in
which the energy space is mapped in the unit interval and $f(E)$ is proportional to a function
$F(s)$ over such interval which parametrically depends on the dimensionless Lagrange multiplier
$b=\beta\phi_0$. Beside being very accurate (typically to order $10^{-10}$), our solution is
complete, in the sense that it holds throughout the unit interval, with the integral transforms
connecting $F(s)$ to $G(s)$, the dimensionless version of the density of states $g(E)$, numerically
computed to high accuracy without any reduction of the integration interval. This is
possible thanks to the reformulation \eqref{eq:pois2} of Poisson's equation, in which the radial
coordinate becomes the unknown to be determined over the span of the gravitational potential (also
mapped to the unit interval). The price to be paid is that only systems with a finite size can be
computed in this way, if high accuracy at any distance is to be preserved.

\smallskip
Our numerical results substantially differ from those reported in \cite{williams2010a}. Lacking any
detailed description of the numerical approach adopted in \cite{williams2010a}, we can only make
some educated guess on the reasons for such manifest discrepancy. First of all, our numerical method
is grounded on the assumption of core density profiles, while the numerical results in
\cite{williams2010a} are claimed to correspond to $1/r$ cusps. It is well possible that DARKexp
admits many different solutions, both cored and cusped. In fact, the shape parameter in
\cite{hjorth2010}, where the DARKexp models was first put forward, is $\beta\Phi(0)$ and in the
numerics of \cite{williams2010a} it takes the values 2.0, 4.0, 2.83, 4.0, 5.66 and 8.0. It is
related to our shape parameter $b$ as
\begin{equation*}
  \beta\Phi(0) =  \Bigl[1+\frac{m}{x(1)}\Bigr]b \,.
\end{equation*}
Since for the second--kind solutions we have $b<b_2=2.0599\ldots$, we see that the two
parametrizations really overlap only when we consider the first--kind solutions, those with small
size. Namely, for the same value of $\beta$, with $\beta\Phi(0)\ge 2$, we have a cored density
profile with rather small size (see Fig.~\ref{fig:rho_slopes}), whereas a density with a much larger
size that oscillates at short distances around the NFW profile is found in
\cite{williams2010a}. Most likely, two different solutions are found because the iterative numerical
method of \cite{binney1982} is used in \cite{williams2010a} starting from the NFW profile itself,
while in our numerical approach we start from a cored, finite size profile.

To clarify this matter, we should try and extend our accurate numerical approach to cusped
systems. However, this extension is not so straightforward if one wants to preserve high accuracy
throughout energy space, since the phase--space distribution of $1/r-$cusped system diverges as
$(E-\phi_0)^{-5/2}$ near the bottom of the potential well. We suspect that this strong divergence,
if not properly treated, might introduce numerical artifacts even when the iterated map happens to
converge. Further work is necessary in this direction.

\smallskip
As we anticipated in the Introduction, we do not try here to apply our findings to realistic
self--gravitating systems. In fact, ergodic phase--space distributions have by themselves a limited
applicability to the outcome of real collapses, where angular momentum plays a crucial role also in
case of perfect spherical symmetry. Our goal in this work, as in the preceding one
\cite{destri2018}, was to develop a framework, both analytic and numeric, for a trustworthy and
accurate solution of the DARKexp problem, since we regard the DARKexp model as a well founded
first--principle approach to the quasi--equilibrium of collisionless self--gravitating matter, which
moreover happens to be compatible with observations and $N-$body simulations.  We have found here a novel
family of ergodic systems with finite size, mass and energy and with a cored density profile. They
match very well the analytic results of \cite{destri2018} and have interesting physical properties
summarized in Figs.~\ref{fig:rho_slopes} and \ref{fig:sigmassq}. We leave for the future the
development of all the extensions and modifications which are necessary to try and apply these ideas to
realistic systems.

\bigskip
\appendix
\section{Accuracy checks and algorithmic details}

We present here an analysis of the accuracy of the numerical results described in section
\ref{sec:results}, providing at the same time more details on the calculations. These were all
performed in MATLAB R2017b with the help of the Chebfun package \cite{chebfun} on a workstation Dell
T7600.

\medskip 
Let us first examine the accuracy of our approach to the integral transforms in eqs.~\eqref{eq:F2nu}
and~\eqref{eq:x2G}. Given a choice of $N$, the two $(N+1)\times(N+1)$ matrices $I^{(F\nu)}$ and
$I^{(xG)}$ [see eqs.\eqref{eq:IFnu}, \eqref{eq:IFnu1} and~\eqref{eq:IxG}] can be computed once and
for all to any desired accuracy with a suitable quadrature tool. We used the {\tt quadgk} routine in
MATLAB with $10^{-12}$ as tolerance, both absolute and relative. Typically, for smooth integrands
such as those in eqs.\eqref{eq:IFnu} and~\eqref{eq:IxG}, the accuracy is much better than the
tolerance, that is to say, the numerical results differ from known analytic ones by much less than
$10^{-12}$.
 
\begin{figure}[tbp]
  \begin{center}
    \includegraphics[width=.7\textwidth]{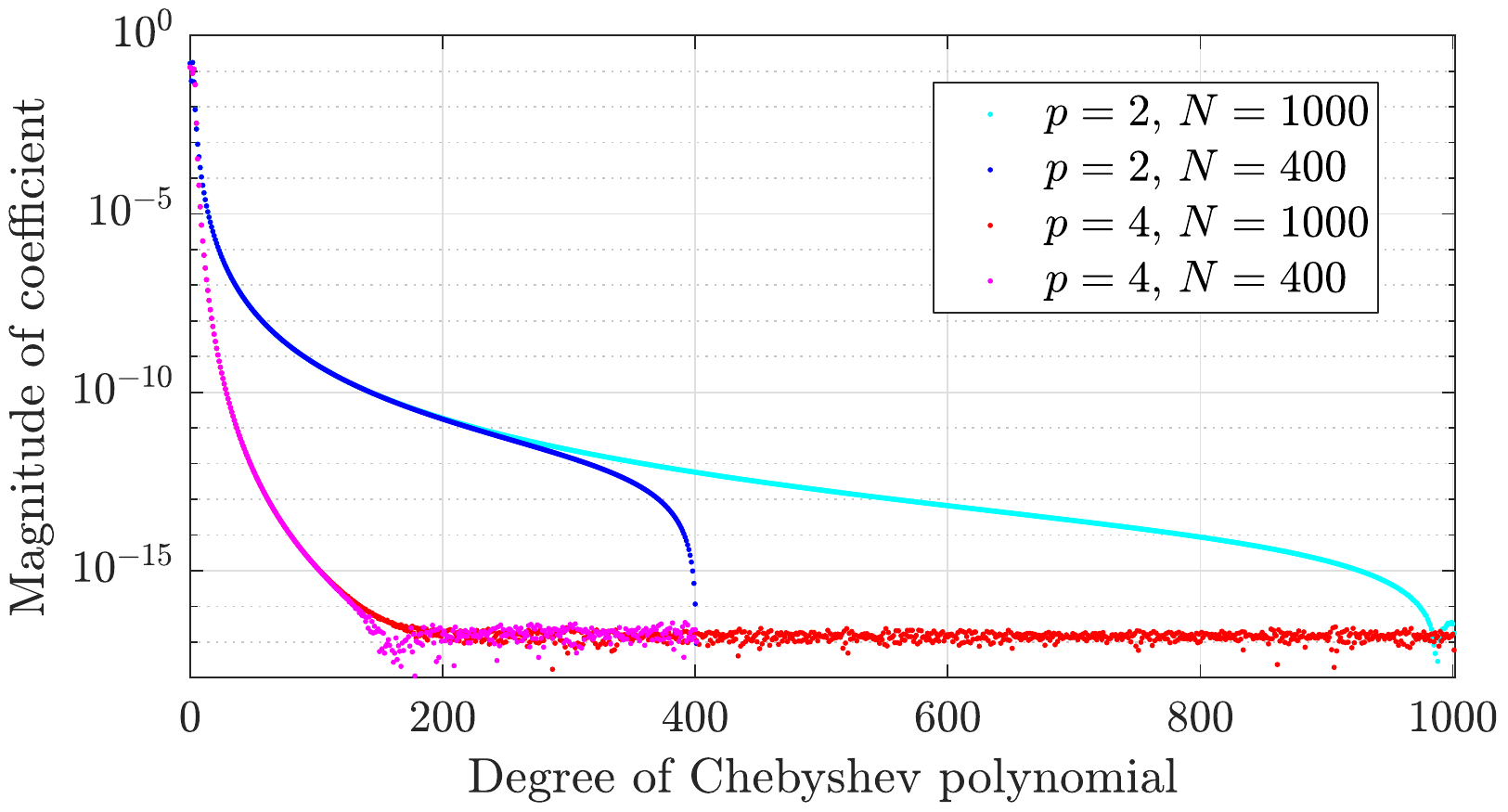}
    \caption{Comparison of the decay of the Chebyshev coefficients of the test function $z^p\log z^p$
      for $p=2$ and $p=4$ when $N=400$ or $N=1000$ Chebyshev point are used. }
    \label{fig:coeffs1}
  \end{center}
\end{figure}

However, the accuracy on the full integrals in eqs.~\eqref{eq:xiF2xinu} and~\eqref{eq:xix2xiG} could
be much worse, and even much larger than $10^{-12}$, depending on the quality of the Chebyshev
approximation \eqref{eq:chebexp} for $\xi_F(z^p)$ and $[\xi_x(z^p)]^3$. This quality is measured by
the decay rate of the corresponding Chebyshev coefficients and can be assessed at the end the
calculation. Given the coefficients, the simplest way to infer the overall accuracy is to test our
quadrature procedure on known test functions replacing $K\xi_F(s)$ and $\xi_x(u)$ in
eqs.~\eqref{eq:xiF2xinu} and~\eqref{eq:xix2xiG}.

The analytic results in eq.~\eqref{eq:nnlasy}, which already suggested the reparametrization
$s=z^p$, with $p$ even to make the leading term $\propto s^{1/2}$ analytic, now suggest also $s\log
s$ as test function. With $p=2$ one obtains first order differentiability in $z=0$ and a sub--optimal
behavior of the Chebyshev coefficients, while with $p=4$ one gets differentiability up to third
order in $z=0$ and very fast convergence  of the coefficients below machine $\epsilon$ as shown in
Fig.~\ref{fig:coeffs1}. The ``exact'' integrations to be used as reference are
\begin{equation*}
  I_1(u)=\tfrac13\int_u^1\!ds\,\frac{s\log s}s\,(s-u)^{1/2} = \tfrac49\left[\tfrac13 (4u-1)(1-u)^{1/2}
    - u^{3/2}\arctan(u^{-1}-1)^{1/2} \right]
\end{equation*}
and
\begin{equation*}
  I_2(s) = \frac8{3\pi s^2}\int_0^s\!du\,(u^{3/2}\log u)^3\,(s-u)^{-1/2} \;.
\end{equation*}
In the case of $I_2(s)$ we do not write nor use the analytic expression, which could nonetheless
be obtained, but rather compute the integral numerically at the grid values of $u$ with the help of
MATLAB's {\tt integral} routine, using $10^{-14}$ as absolute as well as relative tolerance.

We verified that, upon the action of the matrices $I^{(F\nu)}$ and $I^{(xG)}$ over the coefficient
vectors of $z^p\log z^p$ as in eqs.~\eqref{eq:IFnu1} and~\eqref{eq:IxG}, one obtains, if $N=10^3$,
accuracies to order $10^{-14}$ when $p=2$ and even better when $p=4$, practically to the limits of
any computation in double precision. On the other hand, if $N=400$, only with $p=4$ this high
accuracy is preserved, while it gets worse by almost two orders of magnitude when $p=2$. This is still
very accurate, since after all the Chebyshev coefficients of $z^2\log z^2$ with index larger than
$400$ are smaller than $10^{-12}$, but the quality of our approach can only worsen in the true
calculations with $\xi_F(z^p)$ and $[\xi_x(z^p)]^3$. Hence our choice of $p=4$ in the actual
computations whose results are reported in section \ref{sec:results}, with confidence that a
behavior of the Chebyshev coefficients of $\xi_F(z^p)$ and $[\xi_x(z^p)]^3$ similar to those of the
test functions should guarantee a comparable accuracy in the integrations \eqref{eq:xiF2xinu} and
\eqref{eq:xix2xiG}.

\smallskip
The second step $\nu\longrightarrow x$ in the chain \eqref{eq:chain}, which we implement by
numerically solving Poisson's equation in the form eq.~\eqref{eq:pois3}, is by far the most
time--consuming in the chain \eqref{eq:chain}. We used the ODE solvers {\tt ode45} and {\tt ode113}
in MATLAB, with an absolute tolerance of $10^{-15}$ and a relative tolerance ranging from $10^{-6}$
down to $10^{-12}$. The relative tolerance controls in particular the accuracy near the initial
conditions for $\gamma(y)$ and therefore affects the accuracy everywhere else, in particular near
$s=1$. Comparing solutions at different relative tolerances allows to estimate this
accuracy. Another check is possible by comparing the numerical solution to analytic ones, such as
the Schuster model (a.k.a. Plummer's model or $n=5$ polytrope system), which in our dimensionless
setup is written, with $h=1$ in eq.~\eqref{eq:nux2xi},
\begin{equation}\label{eq:I}
  \xi_\nu^{(I)}(u) = (1-u)^5 \;,\quad \xi_x^{(I)}(u) = (1-u/2)^{1/2}\;.
\end{equation}
Another possibility is, still with $h=1$,
\begin{equation}\label{eq:II}
  \xi_\nu^{(II)}(u) = \big(1-u^{1/2}\big)^4 \;,\quad \xi_x^{(II)}(u) = 1+u^{1/2}\;.
\end{equation}
These two analytic solutions of Poisson's equation~\eqref{eq:pois2} have infinite size, with the
dimension--ful density $\rho(r)$ dying off at large $r$ as $r^{-5}$ and $r^{-4}$, respectively.  Our
numerical approach is not set up to deal with infinite size with the necessary accuracy, hence we
restrict the $u-$interval to $[0,a]$ with $a$ slightly less than $1$. By varying $a$ we can check
the accuracy of the numerical solution of eqs.~\eqref{eq:pois2} [in the equivalent form
\eqref{eq:pois3}] by fixing $a$ so that $x^{(I)}(a)\approx x(1)$ or $x^{(II)}(a)\approx x(1)$ for
anyone of the solutions described in section \ref{sec:results}. Notice that in the tests the ODE
solver does not use the known analytic forms of $\xi_\nu^{(I)}(u)$ or $\xi_\nu^{(II)}(u)$, but their
$N-$order Chebyshev interpolations based on the values they assume on the $N$ points defined by
eq.~\eqref{eq:chegrid}.

\begin{table}
\centering
\begin{tabular}{|c|c|c|c|c|c|c|c|}
\hline
{\small $x(1)$} & {\small $a$} & {\small $x^{(I)}(a)$} & {\small $x^{(II)}(a)$} & {\small $\Delta^{(I,400)}$} & {\small $\Delta^{(II,400)}$} & {\small $\Delta^{(I,1000)}$} & {\small $\Delta^{(II,1000)}$} \\
\hline
{\small 5.388} & {\small 0.700} & {\small 3.8006} & {\small 8.6581} & {\small 1.18$\,\cdot10^{-12}$} & {\small 1.96$\,\cdot10^{-12}$} & {\small 8.18$\,\cdot10^{-13}$} & {\small 2.21$\,\cdot10^{-12}$} \\
\hline
{\small 12.703} & {\small 0.875} & {\small 8.4853} & {\small 21.8967} & {\small 1.31$\,\cdot10^{-12}$} & {\small 3.76$\,\cdot10^{-12}$} & {\small 4.28$\,\cdot10^{-12}$} & {\small 2.19$\,\cdot10^{-12}$} \\
\hline
{\small 159.277} & {\small 0.985} & {\small 67.1648} & {\small 187.8520} & {\small 1.96$\,\cdot10^{-11}$} & {\small 1.07$\,\cdot10^{-10}$} & {\small 9.34$\,\cdot10^{-12}$} & {\small 2.42$\,\cdot10^{-10}$} \\
\hline
\end{tabular}
\caption{Accuracy checks in the case of the two analytic solutions of Poisson's equation in
  eqs.~\eqref{eq:I} and~eqs.~\eqref{eq:II}.}
\label{table:table6}
\end{table}

Let us define, for $J=I,II$ and $N=400,1000$, the absolute error of the numerical solution of 
eq.~\eqref{eq:pois3} w.r.t. the exact analytic solution
\begin{equation*}
  \Delta^{(J,N)} = \underset{s}{\rm max}\big|\xi_x^{(J,N)}(s) - \xi_x^{(J)}(s)\big| \;.
\end{equation*}
We find the results listed in Table \ref{table:table6}. As expected, $\Delta^{(J,N)}$ strongly
depends on the scale of $x^{(I)}(a)$ or $x^{(II)}(a)$. We infer that a similar accuracy holds for
the $\xi_x$ of section \ref{sec:results} w.r.t. the unknown exact solution. In particular we see
that when $b>2$ in the family of solutions of the second kind, that is when $x(1) > 159.2771\ldots$,
the overall accuracy of the calculation is entirely determined by the accuracy of the second step
$\nu\longrightarrow x$ in the chain \eqref{eq:chain}, that is by Poisson's equations.

\smallskip 

Let's now come the part of the calculation that drives the integrations and the ODE solver, that is
the solution of eq.~\eqref{eq:xixi}. We have reduced the problem to a system of $N$ non--linear
equations for the $N$ unknowns $\xi_F(z_k^p)$, $k=1,2,\ldots N$, to be dealt with with either one,
or possibly both, of the two approaches described in subsection \ref{sec:nlse}. There is little to
say about the iterative scheme \eqref{eq:simplemap}, which has only one tunable parameter, the
damping factor $\alpha$. The best case if it converges for $\alpha=1$, as it happens for the solutions of
the first kind of subsection \ref{sec:firstkind}. If not, as in the case of the second--kind family
of subsection \ref{sec:secondkind}, one could in principle try few other
values of order one, both positive and negative. However, if the Jacobian matrix $J(1)$ [see
eq.~\eqref{eq:jac}] has eigenvalues with magnitude larger than one, there is a little
chance to gather all eigenvalues of $J(\alpha)$ within the unit circle just by varying
$\alpha$. In fact, this failure just characterizes the second--kind solutions.  In
these circumstances, it is necessary to use algorithms that minimize a suitable objective
function, such as ${\cal F}[\xi_F]$ in eq.~\eqref{eq:f2min}. Through the Optimization Toolbox,
MATLAB offers the specific function {\tt fsolve} for this purpose. As other functions of that
toolbox, {\tt fsolve} has a wealth of options that can be fixed to improve the overall behavior of
the routine. One such option chooses the actual minimization algorithm. We found the
Levenberg--Marquardt algorithm to be very efficient for our problem and used it most of the times.

Besides the precise form of the objective function, which can be changed for instance by varying the
weights $w_k$, one can vary also the parametrization of the step in $\xi_F$. We mostly used the
following, positive--definite and very simple form
\begin{equation*}
  \xi_F(z_k^p) = \xi_F^{(0)}(z_k^p) \exp(y_k)\;, \quad k = 1,2,\ldots,N \;,
\end{equation*}
where $\xi_F^{(0)}$ is the initial configuration and the $y_k$ are the numbers which are varied in
the Levenberg--Marquardt algorithm. Sometime we found it necessary to use also another
parame\-trization, based on the expansion \eqref{eq:asympt}. 

Other important options in {\tt fsolve} are the {\it FunctionTolerance} and the {\it StepTolerance},
which can be fixed to values small enough to obtain the required accuracy when the algorithm
eventually stops, because the objective function does not decrease enough any more ({\it
  FunctionTolerance}) or the proposed change in $\xi_F$ is too small ({\it StepTolerance}). The
final $\xi_F$ is an acceptable solution of eq.~\eqref{eq:xixi} if it satisfies the criterion
\eqref{eq:success}. 

A confirmation that $\epsilon=10^{-11}$ is indeed close to the limits of the double--precision
accuracy for this kind of computations is drawn from the typical profile of the magnitude of the
residuals $R_k$ in a successful run [see Fig.~\ref{fig:iter_resid1}]. This profile is dominated at
small $z$ by the rugged patterns characteristic of roundoff errors. Notice that any numerical error
at low $z$ is inevitably amplified at higher $z$ by the ODE solver in the $\nu\longrightarrow x$
step.  In fact, for more compact systems where $x(u)$ has a smaller variation, we can obtain $|R_k|$
much smaller than $10^{-11}$ and rugged almost throughout the interval $[0,1]$. Another independent
confirmation of the $\epsilon=10^{-11}$ rule comes from the behavior of the maximal residual ${\cal
  R}$ in the iterated map \eqref{eq:simplemap}, as evident from Figs. \ref{fig:simplemap} and
\ref{fig:simplemap3}.

\smallskip 
Quite naturally, there could be also unsuccessful runs of {\tt fsolve}, since the
Levenberg--Marquardt algorithm, as any other local minimization algorithm, may eventually stop at
some $\xi_F$ that is not a solution of eq.~\eqref{eq:xixi}. This could
happen for two reasons: eq.~\eqref{eq:xixi} itself simply has no solution for the specific value of
$b$ one is considering, or $\xi_F^{(0)}$, the starting point passed to the algorithm, is in the
basin of attraction of a local minimum of the objective function which is not its global minimum.
In the latter case the iterative approach \eqref{eq:simplemap} might help to get away from the
local minimum and move toward an acceptable solution. At any rate, as stated near the end of subsection
\ref{sec:nlse}, we cannot consider a failure to produce an acceptable solution through our numerical
methods as a rigorous proof of non--existence of the exact, infinite--precision solution for the given value
of $b$. But it certainly is strong indication of  non--existence.  

\medskip 

The last question concerns the accuracy of an acceptable solution w.r.t. the exact solution. Taking
into account the accuracies in the various steps of the calculation, as discussed above, we are
confident that such as solution is accurate to at least $10^{-10}$, provided the Chebyshev
coefficients of $\xi_F$ and $\xi_x^3$ decay rapidly enough towards zero, in a way similar to that in
Figs.~\ref{fig:coeffs1} or~\ref{fig:coeffs2}. This we verified for all our numerical solutions. The
only exceptions to the $10^{-10}$ accuracy rule are the last four cases of table \ref{table:table4},
because of the worse accuracy in the solution of Poisson's equation. In these cases, an estimate of
the overall accuracy is provided by $\Delta m$ and $\Delta\eta$ in the same table.

\acknowledgments
The author acknowledges the contribution of Michele Turelli in the very early stages of this work.

%\paragraph{Note added.} This is also a good position for notes added
%after the paper has been written.

\end{document}